\documentclass[twocolumn, superscriptaddress,amsmath, prl, aps]{revtex4-2}

\usepackage[T1]{fontenc}
\usepackage[utf8]{inputenc}
\usepackage{siunitx}
\usepackage{times,color,amsthm,graphics,graphicx,bm,bbm,dcolumn}
\usepackage{epsfig}
\usepackage{graphicx}
\usepackage{xcolor}
\usepackage[colorlinks,urlcolor=black,citecolor=blue,linkcolor=black]{hyperref}
\usepackage{soul,bm}
\usepackage{verbatim}
\usepackage{lineno}
\usepackage{orcidlink}

\begin{document}
\setlength{\tabcolsep}{18pt}

\onecolumngrid

\setcounter{equation}{0}
\setcounter{figure}{0}
\setcounter{table}{0}

\clearpage
\setcounter{page}{1}

\title{Mutual friction and vortex Hall angle in a strongly interacting Fermi superfluid}
\date{\today}

\author{N.~Grani~\orcidlink{0000-0001-6107-9726}}
\altaffiliation{These authors contributed equally to this work.}
\email[E-mail: ]{grani@lens.unifi.it}
\affiliation{Department of Physics, University of Florence, 50019 Sesto Fiorentino, Italy}
\affiliation{European Laboratory for Nonlinear Spectroscopy (LENS), University of Florence, 50019 Sesto Fiorentino, Italy}
\affiliation{Istituto Nazionale di Ottica del Consiglio Nazionale delle Ricerche (CNR-INO) c/o LENS, 50019 Sesto Fiorentino, Italy}
\affiliation{INFN, Sezione di Firenze, 50019 Sesto Fiorentino, Italy}

\author{D.~Hern\'andez-Rajkov~\orcidlink{0009-0002-1908-4227}}
\altaffiliation{These authors contributed equally to this work.}
\email[E-mail: ]{grani@lens.unifi.it}
\affiliation{European Laboratory for Nonlinear Spectroscopy (LENS), University of Florence, 50019 Sesto Fiorentino, Italy}
\affiliation{Istituto Nazionale di Ottica del Consiglio Nazionale delle Ricerche (CNR-INO) c/o LENS, 50019 Sesto Fiorentino, Italy}
\affiliation{INFN, Sezione di Firenze, 50019 Sesto Fiorentino, Italy}
\altaffiliation{These authors contributed equally to this work.}

\author{C.~Daix~\orcidlink{0009-0000-7676-8508}}
\altaffiliation[Present address: ]{Laboratoire Kastler Brossel, ENS-Université PSL, CNRS, Sorbonne Université, Collège de France, 24 rue Lhomond, 75005, Paris, France}
\affiliation{Department of Physics, University of Florence, 50019 Sesto Fiorentino, Italy}
\affiliation{European Laboratory for Nonlinear Spectroscopy (LENS), University of Florence, 50019 Sesto Fiorentino, Italy}

\author{P.~Pieri~\orcidlink{0000-0001-8295-805X}}
\affiliation{Department of Physics and Astronomy, University of Bologna, 40126 Bologna, Italy}
\affiliation{INFN, Sezione di Bologna, 40127 Bologna, Italy}

\author{M.~Pini~\orcidlink{0000-0001-5522-5109}}
\affiliation{Institute of Physics, University of Augsburg, 86159 Augsburg, Germany}
\affiliation{Max Planck Institute for the Physics of Complex Systems, 01187 Dresden, Germany}

\author{P.~Magierski~\orcidlink{0000-0001-8769-5017}}
\affiliation{Faculty of Physics, Warsaw University of Technology, 00-662 Warsaw, Poland}
\affiliation{Department of Physics, University of Washington, Seattle, Washington 98195-1560, USA}

\author{G.~Wlazłowski~\orcidlink{0000-0002-7726-5328}}
\affiliation{Faculty of Physics, Warsaw University of Technology, 00-662 Warsaw, Poland}
\affiliation{Department of Physics, University of Washington, Seattle, Washington 98195-1560, USA}

\author{M.~Fr\'ometa Fern\'andez~\orcidlink{0000-0003-4937-8306}}
\affiliation{European Laboratory for Nonlinear Spectroscopy (LENS), University of Florence, 50019 Sesto Fiorentino, Italy}
\affiliation{Istituto Nazionale di Ottica del Consiglio Nazionale delle Ricerche (CNR-INO) c/o LENS, 50019 Sesto Fiorentino, Italy}
\affiliation{INFN, Sezione di Firenze, 50019 Sesto Fiorentino, Italy}

\author{F.~Scazza \orcidlink{0000-0001-5527-1068}}
\affiliation{Department of Physics, University of Trieste, 34127 Trieste, Italy}
\affiliation{Istituto Nazionale di Ottica del Consiglio Nazionale delle Ricerche (CNR-INO), 34149 Trieste, Italy}
\affiliation{European Laboratory for Nonlinear Spectroscopy (LENS), University of Florence, 50019 Sesto Fiorentino, Italy}

\author{G.~Del Pace \orcidlink{0000-0002-0882-2143}}
\affiliation{Department of Physics, University of Florence, 50019 Sesto Fiorentino, Italy}
\affiliation{European Laboratory for Nonlinear Spectroscopy (LENS), University of Florence, 50019 Sesto Fiorentino, Italy}
\affiliation{Istituto Nazionale di Ottica del Consiglio Nazionale delle Ricerche (CNR-INO) c/o LENS, 50019 Sesto Fiorentino, Italy}
\affiliation{INFN, Sezione di Firenze, 50019 Sesto Fiorentino, Italy}

\author{G.~Roati \orcidlink{0000-0001-8749-5621}}
\affiliation{European Laboratory for Nonlinear Spectroscopy (LENS), University of Florence, 50019 Sesto Fiorentino, Italy}
\affiliation{Istituto Nazionale di Ottica del Consiglio Nazionale delle Ricerche (CNR-INO) c/o LENS, 50019 Sesto Fiorentino, Italy}
\affiliation{INFN, Sezione di Firenze, 50019 Sesto Fiorentino, Italy}

\begin{abstract}
The motion of a quantized vortex is intimately connected with its microscopic structure and the elementary excitations of the surrounding fluid. In this work, we investigate the two-dimensional motion of a single vortex orbiting a pinned anti-vortex in a unitary Fermi superfluid at varying temperature. By analyzing its trajectory, we measure the yet-unknown longitudinal and transverse mutual friction coefficients, which quantify the vortex-mediated coupling between the normal and superfluid components. Both coefficients increase while approaching the superfluid transition. They provide access to the vortex Hall angle, which is linked to the relaxation time of the localized quasiparticles occupying Andreev bound states within the vortex core, as well as the intrinsic superfluid parameter associated with the transition from laminar to quantum turbulent flows. We compare our results with numerical simulations and an analytic model originally formulated for superfluid $^3$He in the low-temperature limit, finding good agreement. Our work highlights the interplay between vortex-bound quasiparticles and delocalized thermal excitations in shaping vortex dynamics in unitary Fermi superfluids. Further, it provides a novel testbed for studying out-of-equilibrium vortex matter at finite temperatures.
\end{abstract}

\maketitle

\normalsize
\section{Introduction}

The dynamics of quantized vortices exhibit rich behavior that underlies many phenomena across various quantum fluids. These range from the emergence of quantum turbulence in superfluid helium, atomic and polariton condensates \cite{smith1993decay,tsatsos2016quantum,panico2023onset}, and flux resistance in type-II superconductors \cite{huse1992superconductors}, to the decay of persistent currents \cite{donnelly1998observed, Leggett2006-aw, Halperin2010} and glitches in rotating neutron stars \cite{anderson1975pulsar,zhou2022pulsar,antonelli2020superfluid}. In any finite-temperature superfluid system, as vortices move through the normal component, scattering from thermally-excited quasiparticles gives rise to mutual friction forces \cite{HallVinen}. The microscopic mechanisms governing mutual friction hinge on the intrinsic fundamental properties of the system and its excitations.

Vortex dynamics in fermionic superfluids presents an especially complex problem \cite{kopnin2002vortex}. Excitations above the superfluid ground state include fermionic quasiparticles -- linked to pair-breaking mechanisms -- and bosonic collective modes \cite{Soninbook2015}. In addition, vortices host in-gap Andreev quasiparticles confined inside their cores, which occupy quantized energy levels referred to as Caroli–de Gennes–Matricon (CdGM) states \cite{caroli1964bound,sensarma2006vortices}. 
At temperatures well below the superfluid transition, these bound quasiparticles scatter off delocalized excitations, i.e. the normal fluid. This provides the leading mechanism behind mutual friction in weakly interacting fermionic systems \cite{kopnin2002vortex, Soninbook2015, bevan1997momentum}---well described by Bardeen–Cooper–Schrieffer (BCS) theory. There, the CdGM level spacing $\hbar \omega_0$ is of the order of $|\Delta|^2/E_F$, where $\Delta$ is the superfluid gap and $E_F$ is the Fermi energy, whereas their relaxation time $\tau$ is determined by quasiparticle scattering, depending on the thermal population of normal excitations above the gap. In both ${}^3$He fluids and conventional superconductors where $|\Delta| \ll E_F$, the CdGM spectrum is dense and the quantized nature of CdGM states becomes obfuscated by fast quasiparticle relaxation, especially at high temperature. On the other hand, high-temperature superconductors feature relatively large gaps, $\Delta$ being a significant fraction of $E_F$, and some hints of discrete CdGM levels have been reported \cite{Berthod2017,Chen2018}. 
\begin{figure}[h]
\centering
\vspace{0 pt}
\includegraphics{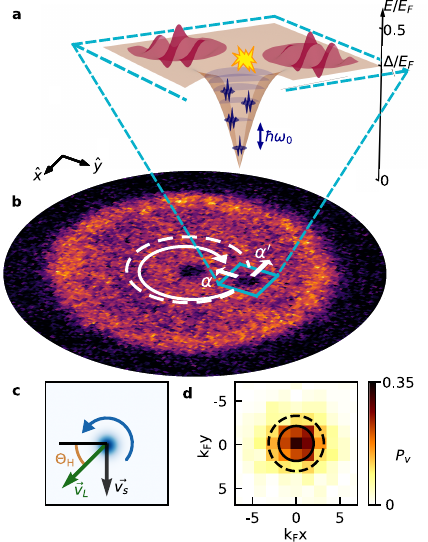}\vspace{-5 pt}
\caption{\textbf{Two-dimensional vortex motion in unitary Fermi superfluids at finite temperature.} 
(\textbf{a}) In Fermi superfluids, the vortex core is populated by localized quasiparticle states (blue wavepackets) characterized by discrete in-gap energies. The energy levels shown in the sketch are calculated using the SLDA for $T = \SI{0.7}{}\,T_c$. 
The scattering process between in-core localized states and delocalized excitations (red wavepackets), is illustrated as a yellow lightning.
(\textbf{b}) Initial configuration of the single vortex dipole. A vortex appears as a depletion in the atomic density in time-of-flight. The effects of the mutual friction coefficients $\alpha$ and $\alpha '$ are visualized in the vortex trajectory (continuous spiraling arrow), deviating from the frictionless circular trajectory (dashed circle). 
(\textbf{c}) Due to mutual friction, the vortex, with circulation direction indicated by the blue arrow, moves with a velocity $\vec{v}_L$, forming an angle $\frac{\pi}{2} - \Theta_H$ with respect to the superfluid velocity $\vec{v}_s$. 
(\textbf{d}) Normalized probability distribution ${P_v}$ of the vortex initial position, characterized by the standard deviations $\sigma_x = \SI{2.55 \pm 0.03}{} k_F^{-1}$ and $\sigma_y = \SI{ 2.06 \pm 0.03 }{}k_F^{-1}$. 
Continuous and dashed circles represent the estimated size of the vortex core $\sim 2k_F^{-1}$, and the imaging resolution $\sim \SI{1}{\mu m}$, respectively.
}
\label{fig:Fig1}
\end{figure}
However, the study of free vortex motion in superconducting materials is hampered by pinning to impurities. Experiments have thus far been mostly focused on superfluid ${}^3$He, where the observed mutual friction has been attributed to the presence of CdGM states~\cite{bevan1997momentum, bevan1997vortex, makinen2018mutual, kopnin1991mutual}.

Advances in ultracold atom experiments have opened new avenues for exploring vortex dynamics in the strongly interacting regime of fermionic superfluidity \cite{ku2014motion,park2018critical,kwon2021sound}. The so-called unitary Fermi gas (UFG) displays the highest critical temperature $T_c$ (normalised to $E_F$) for the superfluid transition of any known fermionic system. It is especially attractive owing to its universal scaling properties, which make it an extraordinary model system for strongly correlated matter. The vortex core size, pair correlations, interparticle spacing, and particle mean free path are indeed governed by a single length scale, namely the inverse Fermi wave vector $k_F^{-1}$. In the UFG, the large energy gap $|\Delta| \simeq 0.47\,E_F$ \cite{novelSF2} leads to $\hbar \omega_0 \sim 0.25\,E_F$, restricting in-core bound states to only a handful of discrete levels (Fig.~\ref{fig:Fig1}a) whose number and thermal population increase with temperature \cite{barresi2023dissipative}. In the presence of such sparse CdGM spectrum, the role of vortex-bound quasiparticle scattering in vortex motion is yet unclear. 

In this work, we investigate the motion of a single vortex in unitary Fermi gases for temperatures well below the superfluid transition. We engineer a minimal two-vortex configuration, i.e.~a vortex dipole, where one vortex orbits around a pinned anti-vortex (Fig.~\ref{fig:Fig1}b). We describe the mobile vortex trajectories by the dissipative point vortex model (DPVM) \cite{schwarz1985three}. Here, the scattering of quasiparticles with the vortex line is phenomenologically accounted for by the longitudinal (dissipative) and transverse (reactive) \textit{mutual friction} coefficients, $\alpha$ and $\alpha'$, respectively \cite{schwarz1985three,kopnin2002vortex,Sergeev2023}. This analytic model successfully describes vortex motion in various systems, including superfluid $^3$He \cite{bevan1997vortex,makinen2018mutual}, $^4$He \cite{minowa2025direct,tang2023imaging}, and atomic superfluids \cite{moon2015thermal,kim2016role,kwon2021sound,neely2024melting}. We extract $\alpha$ and $\alpha'$ as functions of temperature by fitting the vortex trajectories, finding that both increase as the superfluid transition is approached. Our results agree with simulations carried out within a time-dependent density functional theory for superfluid Fermi systems, in the formulation commonly referred to as superfluid local density approximation (SLDA)~\cite{Bulgac2007}. They are also well captured by an analytic model for mutual friction developed for superfluid $^3$He, which accounts for the scattering of bulk excitations with localized quasiparticles occupying Andreev bound states in the vortex core \cite{kopnin2002vortex}. Finally, from the mutual friction coefficients, we determine the vortex Hall angle $\Theta_H$ (Fig.~\ref{fig:Fig1}c) and the intrinsic $q$-parameter of the superfluid. These are important markers controlling the vortex Hall effect in superconducting systems~\cite{heyl2022vortex,ogawa2021large} and the onset of quantum turbulence in neutral superfluids~\cite{finne2003intrinsic}. 

\begin{figure*}[th!]
\centering
\vspace{0 pt}
\includegraphics{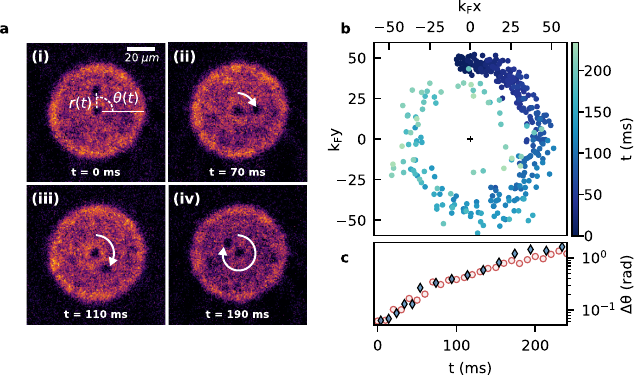}\vspace{-5 pt}
\caption{\textbf{Observation of the orbiting vortex trajectory.}
(\textbf{a}) Imaging individual vortices at different evolution times. Images are averaged over 2-4 experimental realizations. Curved arrows represent the path of the free, orbiting vortex as a function of time. (\textbf{b}) Relative position between the vortices for different realizations as a function of time, encoded by their hue. (\textbf{c}) Time evolution of the standard deviation of the angular position $\mathrm{\Delta \theta}$ for the data in panel (b). Full blue diamonds represent values obtained from the experimental data, empty red circles are obtained from DPVM time evolution. For each evolution time, 40 independent DPVM time evolutions are performed featuring different initial positions of the mobile vortex within its measured confidence range (see Fig.~\ref{fig:Fig1}c).
}
\label{fig:Fig2}
\end{figure*}

\section{Experimental Protocol}

We produce disk-shaped Fermi superfluids with a balanced mixture of the first and third lowest hyperfine states of $\mathrm{^6Li}$ atoms below the critical temperature $T_c$ for the superfluid transition, with $N_{\rm p} = \SI{2.5\pm 0.7}{} \times 10^4$ atoms per spin state. Pair interactions, characterized by the s-wave scattering length $a$, are tuned to the Feshbach resonance located at about $\SI{690}{G}$. We focus on the UFG, defined by the interaction parameter $1/k_F a \sim 0$. Here, $k_F$ is the Fermi wave vector, $k_F = \sqrt{2 m E_F}/\hbar$, determined by the global Fermi energy $E_F$ and the mass  $m$ of a single lithium atom, and $\hbar$ is the reduced Planck's constant. The gas is confined in the $x-y$ plane by a cylindrically symmetric hard-wall repulsive potential of radius $R = \SI{36.5 \pm 0.5}{\mu m}$, realized through a digital micromirror device (DMD).  Along the vertical $z$-direction, the atoms are instead confined within a tight harmonic trap. The homogeneous density in the $x-y$ plane is $n_\mathrm{2D}= \SI{ 6.0 \pm 1.6}{\mu m^{-2}}$ per spin component, while $E_F / h = \SI{8.4 \pm 1.2}{kHz}$ and $k_F =\SI{3.2 \pm 0.2}{ \mu m ^{-1}}$. We control the gas reduced temperature $T/T_F$ (where $T_F = E_F/k_B$ is the Fermi temperature and $k_B$ is the Boltzmann constant) by following different gas preparation procedures. The value of $T/T_F$ is directly measured using a method based on the equation of state similar to \cite{Temp_Measure_Moritz}. See Methods for the sample preparation and temperature determination. We tune $T/T_c$ in the range $0.3 - \SI{0.6}{}$, where $T_c \simeq 0.18\,T_F$ for our trapping geometry, as obtained from a fully self-consistent $t$-matrix approach \cite{Pini2019} (Supplementary Information).

We prepare the initial vortex configuration by using the chopstick method outlined in \cite{Samson_DeterministicCreation, kwon2021sound}. In particular, we create a vortex--anti-vortex pair -- the two-dimensional analogue of a vortex ring\cite{tang2023imaging}-- by translating two $\mu$m-sized optical potentials across the superfluid. Both pinning potentials have a Gaussian profile with a $1/e^2$-width $ \sigma = \SI{1.6 \pm 0.2}{\mu m}$, and intensity $V_0 / E_F = \SI{2.7 \pm 0.5}{}$. The width $\sigma$ is of the same order of the expected vortex core size, $\xi_v \sim 2k_F^{-1} = \SI{0.62\pm0.04}{\mu m}$ \cite{sensarma2006vortices}, and is limited by the resolution of our optical system ($\sim \SI{1}{\mu m}$). One vortex is positioned at the center of the sample, while the other is dragged at a distance $r_0 = \SI{14.3 \pm 0.9}{\mu m}=\SI{45 \pm 6}{} k_F^{-1}$ (Fig.~\ref{fig:Fig1}b). Each vortex carries a single quantum of circulation $\kappa_c = h / 2 m$. We control their initial positions with a precision $ \sim \SI{1.1}\xi_v$, estimated as the standard deviation of the distribution of vortex initial position over repeated realizations (see Fig.~\ref{fig:Fig1}d). This aspect is crucial for a reliable reconstruction of vortex trajectories across many independent measurements.

The vortex dynamics is initialized by removing the potential pinning the off-centered vortex. The free vortex follows a spiraling trajectory, influenced by the background superfluid flow generated by the pinned central vortex and boundary conditions, and mutual friction (Fig.~\ref{fig:Fig2}a-b). The in-plane density homogeneity minimizes the influence of density gradients on vortex motion unlike for harmonically confined gases \cite{moon2015thermal, ku2014motion}. Moreover, the strong confinement along the $z$-direction limits the excitation of Kelvin waves, keeping vortex lines rectilinear, thus making their dynamics effectively two-dimensional (Supplementary Information). During the cyclotron motion of the mobile vortex, only sound waves having wavelengths $ \lambda \gg r_0$ are expected to radiate out of the accelerating vortex \cite{vinen2000classical}, negligibly affecting the observed trajectories. 

We track vortex trajectories over hundreds of milliseconds (Fig.~\ref{fig:Fig2}b), performing at least 30 repetitions for each evolution time. Despite the precise initial positioning of the vortices, the angular standard deviation $\Delta \theta$ increases with the evolution time (Fig.~\ref{fig:Fig2}c). We study this behavior, by analyzing the time evolution of multiple vortex trajectories with the DPVM model, starting from slightly different initial positions, accounting for the experimental uncertainty.
The observed agreement between the experimental and numerical evolution of $\Delta\theta(t)$ suggests that the observed spread in the data is fundamentally limited by the intrinsic constraints of manipulating quantum vortices with finite resolution, even though comparable to the core size. 

\begin{figure*}[ht!]
\centering
\vspace{0 pt}
\includegraphics{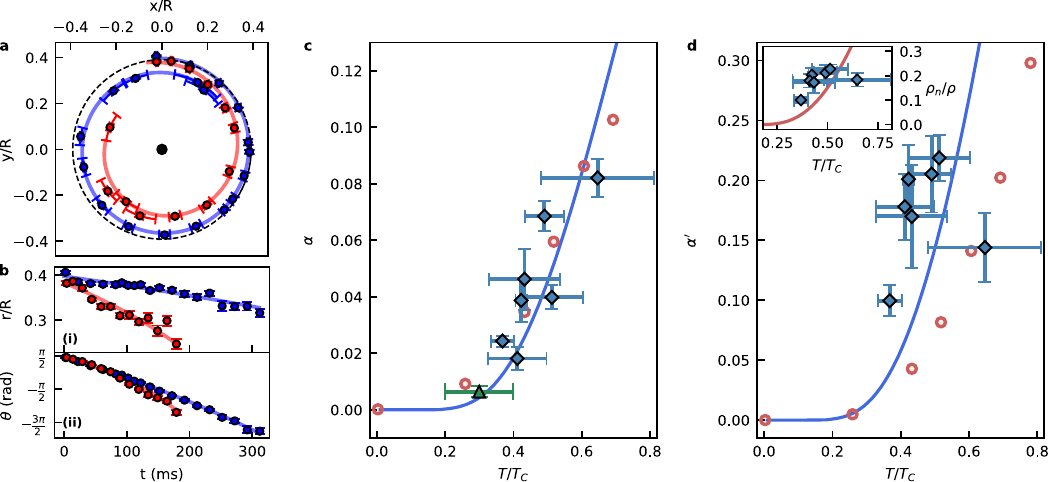}\vspace{-5 pt}
\caption{\textbf{Temperature dependence of the mutual friction coefficients.}
Observed vortex trajectories (\textbf{a}) decomposed on the radial [\textbf{(\textbf{b})-i}] and azimuthal [\textbf{(\textbf{b})-ii}] coordinates over time for different temperatures: $T/T_c= \SI{0.36 \pm 0.04}{}$ (blue) and $T/T_c= \SI{0.50 \pm 0.06}{}$ (red). Data points represent the average angular and radial position of the vortex, error bars are standard deviation of the mean. Solid curves represent fits of the experimental trajectories using the DPVM. The dashed circumference shows the frictionless vortex trajectory. Measured mutual friction coefficients: (\textbf{c}) $\alpha$ and (\textbf{d}) $\alpha'$ as a function of temperature.  Experimental data are shown as blue diamonds, vertical error bars correspond to the fitting error of the trajectories, horizontal error bars correspond to the uncertainty in the temperature determination. The green triangle shows the value of $\alpha$ obtained in Ref.~\cite{kwon2021sound}. Red open circles are obtained from the time-dependent SLDA simulations for $k_F r_0 = 31$. The blue curves corresponds to Eq.~\eqref{alphakopnin} and \eqref{alpha_prime_kopnin} using $\rho_s/\rho$,  $\Delta/\Delta(T=0)$, and $\hbar\omega_0=|\Delta|^2/E_F$ from SLDA, and $T_c\simeq 0.17\,T_F$ (homogeneous case). The inset shows the estimated normal component extracted from Eq.~\eqref{alpha_prime_kopnin}, with the solid red line representing $\rho_n/\rho$ from SLDA.
}
\label{fig:Fig3}
\end{figure*}

\section{Temperature dependence of the mutual friction}

We investigate the effects of thermal excitations on vortex dynamics by performing measurements at different temperatures, within the range $0.3 - \SI{0.6}{}\,T_c$. In Fig.~\ref{fig:Fig3}a-b, we present two typical experimental trajectories for $T/T_c= \SI{0.36 \pm 0.04}{}$ and $T/T_c= \SI{0.50 \pm 0.06}{}$. In general, we observe that at higher temperatures the trajectories exhibit more pronounced deviations from circular motion. To extract the mutual friction coefficients, we fit the experimental trajectories using the equation of motion of the mobile vortex of the DPVM, assuming the normal component remains stationary in the laboratory frame \cite{moon2015thermal, simjanovski2024shear, neely2024melting} (see Methods). The values of $\alpha$ are obtained from the time-evolution of the radial coordinate $r$, while those of $\alpha'$ are extracted from the temporal lag of the azimuthal coordinate $\theta$. Both coefficients increase with temperature (Fig.~\ref{fig:Fig3}c-d), reflecting the increasing normal component. Interestingly, $\alpha'$ remains finite over the entire temperature range explored, in contrast to previous experiments and theories for atomic bosonic superfluids, where it was considered to be negligible~\cite{Jackson2009,Sergeev2023}. At the same time, the measured values of $\alpha$ are much higher than those reported for a weakly-interacting BEC over a comparable temperature range, where the maximum value was found to be $\alpha \simeq 0.03$ at $T\simeq0.8\,T_c$ \cite{moon2015thermal}. In that case, the dominant contribution to mutual friction was believed to arise from the scattering of collective excitations (phonons) by the vortex velocity field. The higher sound speed in Fermi superfluids~\cite{novelSF2}, instead, leads to an energetic suppression of Bogoliubov-Anderson phonon excitations, potentially reducing their role in mutual friction. However, their contribution is yet to be theoretically determined ~\cite{Mozyrsky2019, pitaevskii1959calculation}. Furthermore, the measured dissipative coefficient $\alpha$ exhibits an intermediate trend between those measured in ${}^3$He \cite{bevan1997momentum} and ${}^4$He \cite{barenghi1983friction} fluids. This behavior fits in with the values measured at low temperatures when moving from the BEC towards the BCS regime \cite{kwon2021sound}, being compatible with an increasing number of CdGM states due to the exponential gap reduction in BCS superfluids.

We carry out numerical simulations using the time-dependent SLDA, a fully microscopic approach that is formally equivalent to the mean-field Bogoliubov–de Gennes equations for Fermi systems (see Methods). This method has been previously shown to provide a quantitative description of dissipative vortex dynamics in Fermi superfluids \cite{Bulgac2007, barresi2023dissipative}. We apply the same DPVM model fitting protocol to analyze the SLDA-simulated vortex trajectories, extracting predictions for $\alpha$ and $\alpha'$ shown in Fig.~\ref{fig:Fig3}c-d. The SLDA simulations accurately reproduce the experimental values of $\alpha$, while capturing the qualitative trend of $\alpha'$. As expected, at low temperatures, both coefficients approach zero, reflecting the vanishing population of the normal component. Similar to the experimental case, for $T > 0.3\,T_c$, $\alpha'$ remains finite and exceeds $\alpha$.

\begin{figure}[t]
\centering
\vspace{0 pt}
\includegraphics{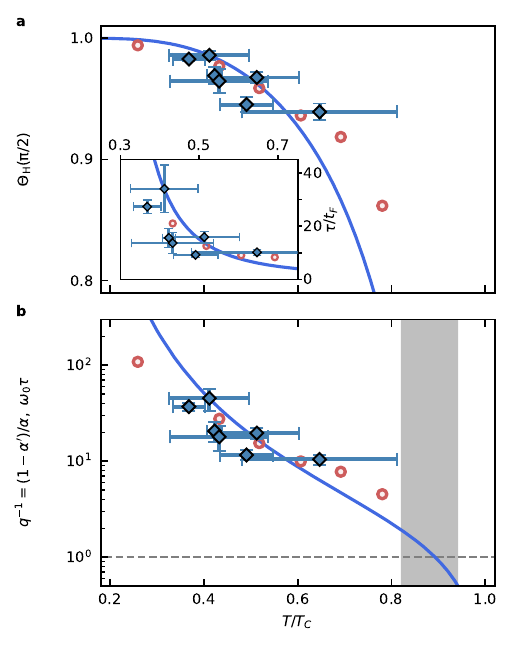}\vspace{-5 pt}
\caption{\textbf{Vortex Hall angle and intrinsic superfluid $q$-parameter.}
(\textbf{a}) The vortex Hall angle $\Theta_H = \tan^{-1}\left((1-\alpha')/\alpha\right)$ as a function of temperature, determining the deviation of the vortex velocity from the normal to the background superfluid velocity. The inset shows the estimated relaxation time $\tau$ of localized quasiparticle, normalized to the Fermi time $t_F = h/E_F = \SI{0.12 \pm 0.02}{ms}$, determined from $\tan \Theta_H=\omega_0\tau$ by fixing $\hbar \omega_0 = |\Delta|^2 / E_F$. (\textbf{b}) Intrinsic parameter $q^{-1} = (1-\alpha') / \alpha$ as a function of temperature. The shaded gray region defines the estimated transition temperature to the laminar regime, $q \sim 1$. Specifically, it marks the regime of temperatures providing a value of $q^{-1}$ between 0.5 and 2 according to Kopnin formula (blue line). Color code is defined in Fig.~\ref{fig:Fig3}c-d.
}
\label{fig:Fig4}
\end{figure}

To connect our results with the microscopic mechanisms underlying the mutual friction in fermionic superfluids, we compare them with the analytical predictions developed by Kopnin for describing mutual friction in $^3$He and weakly coupled BCS systems \cite{kopnin2002vortex, kopnin1991mutual}. In this approach, mutual friction at low temperatures originates from the scattering between quasiparticle excitations above the gap and CdGM states, while the scattering of quasiparticles with the vortex velocity field are neglected \cite{kopnin1991mutual}. Additionally, it incorporates the Iordanskii force arising from the Aharonov-Bohm effect of quasiparticles scattering with the topological vortex phase structure \cite{Soninbook2015, kopnin1991mutual}.
The mutual friction coefficients take the form:
\begin{align}
    \alpha  &= \frac{\rho_s}{\rho}\left(\omega_0\tau\tanh\frac{|\Delta|}{2k_BT}\right)^{-1} ,
\label{alphakopnin} \\
    \alpha' &= 1- \frac{\rho_s}{\rho}\left(\tanh\frac{|\Delta|}{2k_BT}\right)^{-1},
\label{alpha_prime_kopnin}
\end{align}
where $\rho =\rho_s+\rho_n$ is the total density, $\rho_s/\rho$ is the superfluid fraction, and $\tau \propto \frac{\rho}{\rho_n}\tau_n$, with $\tau_n \sim E_F/T_c^2$ being the Fermi-liquid relaxation rate in the normal state at the critical temperature \cite{kopnin2002vortex, kopnin1991mutual} (see Methods). In particular, the dissipative parameter $\alpha$ is proportional to the scattering rate $\Gamma \sim \tau^{-1}$ of quasiparticles confined within the vortex core with delocalized excitations, compatible with an increasing momentum exchange between the normal and superfluid components as scattering events become more frequent. Conversely, the reactive $\alpha '$ depends only on the properties of the bulk superfluid. 

Remarkably, as shown in Fig.~\ref{fig:Fig3}c-d, this analytic approach closely reproduces the observed behavior of both $\alpha$ and $\alpha '$. The agreement between the experimental data and Kopnin's prediction suggests that CdGM states play a key role in the emergence of mutual friction in unitary Fermi superfluids. The remaining discrepancy may stem from the finite number and nonlinear spectrum of CdGM levels at unitarity, deviating from the weakly interacting BCS limit considered within Kopnin's model (Supplementary Information). Additional contributions from in-bound scattering events within the vortex core could further enhance mutual friction \cite{silaev2012universal}, yielding small yet non-zero dissipation even for $T\rightarrow0$~\cite{eltsov2007}. Notably, Kopnin's model quantitatively reproduces also the behavior observed in SLDA simulations (see Extended~Data~Fig.~\ref{fig:extFigds} in Methods). The deviation seen for $\alpha'$ in Fig.~\ref{fig:Fig3} can be traced back to the overestimation of $T_c \simeq 0.3 \ T_F$ in the SLDA, compared to the experimentally determined value of $T_c = 0.17(1)\ T_F$ for a homogeneous system \cite{novelSF2}. From Eq.~\eqref{alpha_prime_kopnin} we estimate the normal fraction $\rho_n/\rho$, and find it follows a consistently increasing behavior with temperature. We find somewhat higher values with respect to recent experimental results~\cite{yan2024thermography}. However, we remark that measuring $\rho_n$ at low temperatures remains a challenge \cite{li2022second}.

The relative strength between transverse and longitudinal components of the mutual friction force offers a clear geometric description of vortex motion, which mirrors charge transport in electric and magnetic fields. In ordinary metals, deviations from a pure cyclotron orbit are characterized by the Hall angle, proportional to the ratio between the transverse (Hall--reactive) and the longitudinal (Ohmic--dissipative) conductivities. Analogously, the vortex Hall angle $\Theta_H = \tan^{-1}\left((1-\alpha')/\alpha\right)$ \cite{ogawa2021large, heyl2022vortex} describes how a vortex moves with respect to the background superfluid velocity field $\vec{v}_s$. As temperature increases and mutual friction becomes more significant, the direction of the vortex velocity $\vec{v}_L$ shifts relative to $\vec{v}_s$. This results in $\Theta_H$ deviating from $\pi/2$, its value in the absence of dissipation, corresponding to the vortex moving together with the superfluid velocity (see Fig.~\ref{fig:Fig4}a and Fig.~\ref{fig:Fig1}c). The measurement of $\Theta_H$ provides important insights into the properties of quasiparticles localized within the vortex core. In Kopnin's approximation where the mutual friction is entirely determined by the interaction of vortex bound states with delocalized quasiparticles above gap, the vortex Hall angle is directly related to the scattering time of the localized quasiparticles $\tau$ and $\hbar \omega_0$ through the relation $\tan \Theta_H = \omega_0 \tau$. Fixing $\hbar \omega_0 = |\Delta|^2 / E_F$, we can obtain a reasonable estimate of $\tau$ (Fig.~\ref{fig:Fig4}a) \cite{kopnin1991mutual}. Interestingly, we observe that $\tau$ increases with decreasing temperature, while remaining significantly shorter than the timescale of vortex dynamics. This ensures that the observed trajectories reflect a steady equilibrium of scattering events. In addition, the measured values of $\omega_0 \tau$ suggest that, for unitary superfluids, the broadening of the CdGM energy levels, $\delta E_{\rm CdGM} \sim \tau^{-1}$, is small with respect to level separation $\sim \hbar\omega_0$ in the explored temperature range. Analogously to \emph{ultraclean} superconductors \cite{ogawa2021large}, these results open prospects for directly probing discrete CdGM states using spectroscopic techniques.

Even though our experiments are performed in a system containing two vortices, they provide an inroad into regimes of superfluid turbulence, through the intrinsic superfluid $q$-parameter defined as $q=\alpha/(1-\alpha')$. This velocity-independent marker is fundamental to characterize the transition from laminar to turbulent flow in superfluids \cite{finne2003intrinsic,tsatsos2016quantum}, as it parametrizes the crossover from Kelvin waves free to propagate along the axis of a vortex filament ($q^{-1} > 1$) to overdamped Kelvin waves ($q^{-1} < 1$) \cite{barenghi1985thermal, finne2003intrinsic, volovik2003classical, minowa2025direct,makinen2023rotating}. The damping of these excitations is tightly connected to the rate of momentum exchange between the normal and superfluid components, owing to $q^{-1}=\omega_0\tau$. For slow rates, $\Gamma \ll \omega_0$ or $q^{-1} \gg  1$, the system is unable to dissipate vortex excitations, cascading into chaotic vortex dynamics, and allowing for quantum turbulence to develop. On the other hand, when the relaxation is efficient, $\Gamma \gtrsim \omega_0$ or $q^{-1}\lesssim 1$, Kelvin waves relax keeping vortices rectilinear, resulting in laminar flow. The description of quantum turbulence relies additionally on the (extrinsic) superfluid Reynolds number, $\text{Re}_s = uL/\kappa_c$, where $u$ and $L$ denote the characteristic velocity and length scales of the superflow \cite{finne2003intrinsic, volovik2003classical,tsatsos2016quantum}. As long as $q^{-1}\gtrsim1$, the superflow becomes turbulent only for $\text{Re}_s\gg1$, while it remains laminar whenever $\text{Re}_s \sim 1$ \cite{finne2003intrinsic, volovik2003classical}. Within the temperature range explored in this work, we find $q^{-1} \sim 10-40$ (Fig.~\ref{fig:Fig4}b). Thus, although our two-vortex system lies well inside the laminar regime with $\text{Re}_s \simeq 1$ ($u\simeq\kappa_c/r$ and $L\simeq r$), our results suggest that the UFG is suitable to the study of quantum turbulence in many-vortex systems featuring $\text{Re}_s \gg 1$~\cite{slda-turbulence-2022,slda-turbulence-2024}. We estimate that the transition to a regime where the flow is laminar independently on the vortex number occurs at a temperature near the superfluid transition, $T \sim \SI{0.9}{}\,T_c$, where $q^{-1} \sim 1$. This is an intermediate scenario compared with $^4$He and $^3$He superfluids \cite{finne2003intrinsic,donnelly1991quantized,eltsov2007, Sergeev2023}. In $^3$He-B (at $\SI{10}{bar}$) this crossover is observed for lower values of $T/T_c$ \cite{bevan1997momentum}, due to the much larger value of $\alpha$---an outcome of the much denser spectrum of CdGM bound states, $\hbar \omega_0 \sim 10^{-4}E_F$.

\section{Conclusions and Outlooks}

In summary, we have measured the longitudinal and transverse  mutual friction coefficients in the unitary Fermi superfluid at finite temperatures. By comparing our results with numerical SLDA simulations and analytical predictions developed for BCS superfluids, we highlight the essential role of localized Andreev bound states in the dissipative dynamics of quantum vortices. The large Hall angles observed across the investigated temperature range suggest that the UFG operates in the \textit{ultraclean}-core limit, where the discrete nature of core-bound states becomes pronounced. In this regime, which has been difficult to achieve in other Fermi systems \cite{kopnin2002vortex,ogawa2021large}, vortices exhibit low-viscosity dynamics. 

In the future, we plan to investigate other superfluid regimes within the BEC-BCS crossover. Exploring the BEC regime will provide quantitative insights into mutual friction mechanisms that remain uncharacterized, particularly concerning $\alpha'$, linking to the unsolved problem surrounding the Iordanskii force in bosonic superfluids \cite{Sergeev2023}. Conversely, in the BCS regime, CdGM states are expected to have an even stronger influence \cite{barresi2023dissipative, simonucci2019bound}, leading to enhanced dissipation \cite{kwon2021sound, bevan1997vortex} and the emergence of a non-negligible vortex inertial mass \cite{richaud2024dynamical}. The ability to precisely track vortex motion will allow the study of disordered phase structures in BCS spin-imbalanced systems \cite{magierski2024quantum}. Moreover, the latter scenario represents a unique system offering key insights into the polarization dependence of mutual friction coefficients. This dependence stems from modifications of the CdGM spectrum in spin-imbalanced vortex cores \cite{magierski2022reverse}. Finally, our single-vortex pinning protocol paves the way for highly controlled investigations of the mechanisms behind vortex pinning \cite{blatter1994vortices}, particularly its stability against thermal fluctuations \cite{fisher1991thermal} and sound waves \cite{patel2020universal, yan2024thermography}. In the limit of many vortices, this will enable the study of vortex depinning avalanches, which are thought to underlie neutron stars glitches \cite{liu2024vortex}.

\section{Acknowledgements}

We thank B.~Haskell, M.~Antonelli, and S.~Autti for the discussions. We thank M.~Inguscio, J.~Makinen and D. Galli for careful reading of the manuscript, and  W.~J.~Kwon for participating in the initial setting of the experiment. G.R., G.D.P., and P.P.~acknowledge financial support from the PNRR MUR project PE0000023-NQSTI. G.R. acknowledges funding from the Italian Ministry of University and Research under the PRIN2017 project CEnTraL and the Project CNR-FOE-LENS-2024. The authors acknowledge funding from INFN through the RELAQS project. The authors acknowledge support from the European Union - NextGenerationEU for the “Integrated Infrastructure initiative in Photonics and Quantum Sciences" - I-PHOQS [IR0000016, ID D2B8D520, CUP B53C22001750006]. This publication has received funding under the Horizon Europe program HORIZON-CL4-2022-QUANTUM-02-SGA via project 101113690 (PASQuanS2.1) and by the European Community's Horizon 2020 research and innovation program under grant agreement n° 871124. This work was financially supported by the (Polish) National Science Center Grants No. 2022/45/B/ST2/00358 (GW) and 2021/43/B/ST2/01191 (PM). P.P.~ acknowledges funding from the Italian Ministry of University and Research (MUR) under project PRIN2022, Contract No.~2022523NA7 and from the European Union – Next Generation EU through MUR project PNRR - M4C2 - I1.4 Contract No.~CN00000013. F.S. acknowledges funding from the European Research Council (ERC) under the European Union’s Horizon 2020 research and innovation programme (Grant agreement No. 949438).
We acknowledge Polish high-performance computing infrastructure PLGrid for awarding this project access to the LUMI supercomputer, owned by the EuroHPC Joint Undertaking, hosted by CSC (Finland) and the LUMI consortium through PLL/2024/07/017603.

%

\clearpage

\renewcommand{\figurename}{Extended Data Fig.}
\setcounter{figure}{0}

\section*{Methods}
\subsection*{Superfluid sample preparation}
We prepare the unitary superfluid by evaporating a balanced mixture of the hyperfine states $|1\rangle=|F,m_F\rangle= |1/2,1/2\rangle$ and $|3\rangle=|F,m_F\rangle= |3/2,-3/2\rangle$ of $^6$Li, near their scattering Feshbach resonance at $\SI{690}{G}$ \cite{zurn2013precise} in an elongated, elliptic optical dipole trap. A repulsive $\mathrm{TEM_{01}}$-like optical potential at $\SI{532}{nm}$ is then adiabatically ramped up in $\SI{100}{ms}$ before the end of the evaporation to provide strong vertical confinement, with trapping frequency $\omega_z = 2\pi \times 560(5)$ or $640(5)$ Hz. Successively, in the $x$–$y$ plane, a box-like potential is turned on to trap the resulting sample in a circular region. This circular box is projected using a Digital Micromirror Device (DMD). When both potentials have reached their final configuration, the infrared lasers forming the crossed dipole trap are adiabatically turned off, completing the transfer into the final pancake trap. A residual radial harmonic potential of $\SI{2.5}{Hz}$ is present due to the combined effect of an anti-confinement provided by the $\mathrm{TEM_{01}}$ laser beam in the horizontal plane and the confining curvature of the magnetic field used to tune the Feshbach field. This weak confinement has a negligible effect on the sample over the $R=\SI{36.5 \pm 0.5}{\mu m}$ radius of our box trap, resulting in an essentially homogeneous density. We estimate the Fermi energy as \cite{hernandez2024connecting}:
\begin{equation}
    \frac{E_F}{h}=\frac{1}{\pi} \left(\frac{\hbar}{m R^2}  \omega_z N_{\rm p} \right)^{1/2}.
\end{equation}

\begin{figure*}[ht!]
\centering
\vspace{0 pt}
\includegraphics{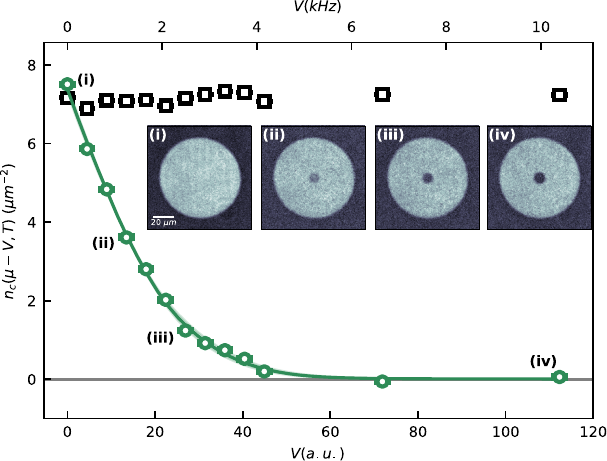}\vspace{-5 pt}
\caption{\textbf{Thermometry protocol.} Measured density in the region where the optical potential is applied as a function of its intensity (green points), and fitted curve using Eq.~\eqref{eq:EOS_Fit} (continuous line). The insets show the observed profile at different values of the external potential. The empty black squares indicate the observed density outside the applied potential.
}
\label{fig:extFig1}
\end{figure*}

\subsection*{Temperature estimation}

To measure the temperature of our system, we adopt a similar protocol as in \cite{Temp_Measure_Moritz}. We probe the equation of state of the UFG by shining an optically homogeneous potential at the center of the cloud with the DMD (see Extended~Data~Fig.~\ref{fig:extFig1}). 

The vertical confinement provided by the $\mathrm{TEM_{01}}$ laser beam is such that $E_F/\hbar \omega_z \approx 15$, making the system collisionally three dimensional. This allows us to use the 3D density distribution of a unitary Fermi gas written in the local density approximation:
\begin{equation}
n_{3D}(\vec{r};\mu,T) = \frac{1}{\lambda_{DB}^3} f_n\left[\beta(\mu-V(\vec{r}))\right],
\end{equation}
where $\beta=1/(k_B T)$ and $\lambda_{DB}=\sqrt{2\pi \hbar^2/ mk_B T}$ is the thermal de Broglie wavelength, $m$ the mass of a $^6$Li atom, $V(\vec{r}) = \frac{1}{2} m\omega_z^2 z^2 + U_{box}(r, \theta)$ the confining potential, and the equation of state $f_n(q)$ defined as \cite{del2021tunneling}:
\begin{equation}
f_n(q) = \left\{
     \begin{array}{lr}
       \sum_{k=1}^4  k\ b_k\ e^{kq}& q<-0.9\\
       -\mathrm{Li}_{3/2}(-e^q)F(q) & -0.9<q< 3.9 \\
       \frac{4}{3\sqrt{\pi}} \left[\left(\frac{q}{\xi}\right)^{\frac{3}{2}} - \frac{\pi^4}{480} \left(\frac{3}{q}\right)^{\frac{5}{2}}\right]& q>3.9\\
     \end{array}
   \right.
\end{equation}
where $F(q)=n(q)/n_0(q)$ is the ratio between the unitary and the non-interacting Fermi gas density measured experimentally in \cite{ku2012revealing} and Li$_{3/2}(x)$ is the polylogarithm of order $3/2$ and argument $x$. The column density measured during the imaging process is:
\begin{equation}\label{eq:EOS_Fit}
n_{c}(\mu,T) = \frac{1}{\lambda_{DB}^3}\int_{-\infty}^{\infty} f_n \left(\beta\mu- \frac{1}{2} m \beta \omega_z^2 z^2\right) d z.
\end{equation}

To extract the temperature and chemical potential, we fit the measured column density $n_{c}(\mu-V,T)$ in the region where the optical potential is applied using Eq.~\eqref{eq:EOS_Fit} for different values of $V$, where $V$ is the height of the barrier. The height of the step is tuned by changing the arrangement of the mirrors of the DMD to achieve different reflection intensities, $I$, which translated to the optical potential $V = \zeta I$, where $\zeta$ is a calibration constant. To obtain the ratio $T/T_F$, we use the fitting result of Eq.~\eqref{eq:EOS_Fit} for the temperature $T$ and the expected Fermi temperature $T_F$ given by the atomic density. Moreover, the chemical potential $\mu$ and the calibration constant $\zeta$ are obtained by performing the fitting with Eq.~\eqref{eq:EOS_Fit} similarly to \cite{Temp_Measure_Moritz}. We note that the value of the calibration constant is comparable to that measured following the method highlighted in \cite{kwon2020strongly}. 

As shown by the black points in Extended~Data~Fig.~\ref{fig:extFig1}, the outer density is not affected by the application of the external potential, confirming that the cloud is not dramatically perturbed by its application. For most of the experimental realization, the thermometry procedure was performed before and after the vortex dynamics experiment. In this case, the temperature value is determined as the average and maximum deviation between the results of the two independent measures. We tune the value of $T/T_F$ temperature of the cloud by transferring the atoms into the final trap with varying depths, at the end of the evaporation process and loading the atoms in the final configuration using a lower or higher initial trapping potential, subsequently set to the values of interest.

\subsection*{Dissipative Point Vortex Model}

The effect of dissipation on the vortex dynamics can be introduced in the context of the two-fluid model as the effect induced by mutual friction within the normal and superfluid components \cite{Sonin_magnusForce}. According to the two-fluid model, the total density is $\rho = \rho_s + \rho_n$, where $\rho_s$ and $\rho_n$ are the density of the superfluid and normal components, respectively. In the description of the dynamics of the vortex, with $\vec{v}_s$ and $\vec{v}_n$ we refer to the background superfluid and normal velocities, namely the velocities of the superfluid and normal components, respectively, in the absence of the vortex.

In our configuration, the motion of the vortex is expected to be determined by the mutual friction force per unit length $\vec{F}_N$ in the superfluid region in proximity to the vortex core, caused by scattering of thermal quasiparticles. In a homogeneous system, for a vortex moving in the $\hat{x}- \hat{y}$ plane, this can be written in the form \cite{Sonin_magnusForce,kopnin2002vortex}:
\begin{equation} 
\label{normal_force}
\vec{F}_N = D(\vec{v}_n - \vec{v}_L) + D' \hat{z}\times (\vec{v}_n - \vec{v}_L),
\end{equation}
where $\vec{v}_L$ is the vortex velocity and $\hat z$ is the unit vector orthogonal to the plane of vortex motion.
The dynamics of a vortex sets a change of momentum in the superfluid flow. For massless vortices, the momentum balance sets \cite{Sonin_magnusForce}:
\begin{equation} 
\label{eq:MomentumBalance}
\vec{F}_N + \vec{F}_M = 0,
\end{equation}
where $\vec{F}_M = \kappa \rho_s (\vec{v}_s - \vec{v}_L) \times \hat{z}$ is the Magnus force.

From Eq.~\eqref{eq:MomentumBalance} it is possible to write the vortex velocity in terms of two mutual friction coefficients $\alpha$ and $\alpha'$, associated with dissipative and reactive terms, respectively~\cite{kopnin2002vortex}. For two-dimensional vortex dynamics, the velocity of the vortex $\vec{v}_L = \frac{d \vec{r}_L}{dt}$ moving in the $\hat{x}-\hat{y}$ plane is described by:
\begin{equation}
\label{VortexVelocity}
\frac{d \vec{r}_L}{dt} = \vec{v}_s + \alpha'(\vec{v}_n  - \vec{v}_s) + \alpha \sigma\hat z \times (\vec{v}_n  - \vec{v}_s).
\end{equation}
where $\sigma= \pm 1$ is the circulation sign ($\sigma \kappa_c = \sigma  h / 2 m$). The coefficients $\alpha$ and $\alpha '$ are related to $D$ and $D'$ by \cite{kopnin2002vortex}:
\begin{align}
\label{eq:alpha}
\alpha &= \frac{d_{||}}{d_{||}^2 + (1-d_{\perp})^2},\\   
\label{eq:alphaprime}
\alpha' &= 1-\frac{1-d_{\perp}}{d_{||}^2 + (1-d_{\perp})^2},\\
\label{eq:d_s}
d_{||} &= \frac{D}{\kappa_c\rho_s}, \quad  \quad \quad d_{\perp} = \frac{D'}{\kappa_c\rho_s}.
\end{align}
To take into account the effects of the boundaries, we consider the contribution of imaginary vortices located at $\tilde{\vec{r}}_L = \frac{R^2}{|r_L|^2} \vec{r}_L$ for the off-center vortex \cite{Imaginary_vortex_Disk}. The velocity field of the central vortex fulfills the boundary condition.

\begin{figure*}[th!]
\centering
\vspace{0 pt}
\includegraphics{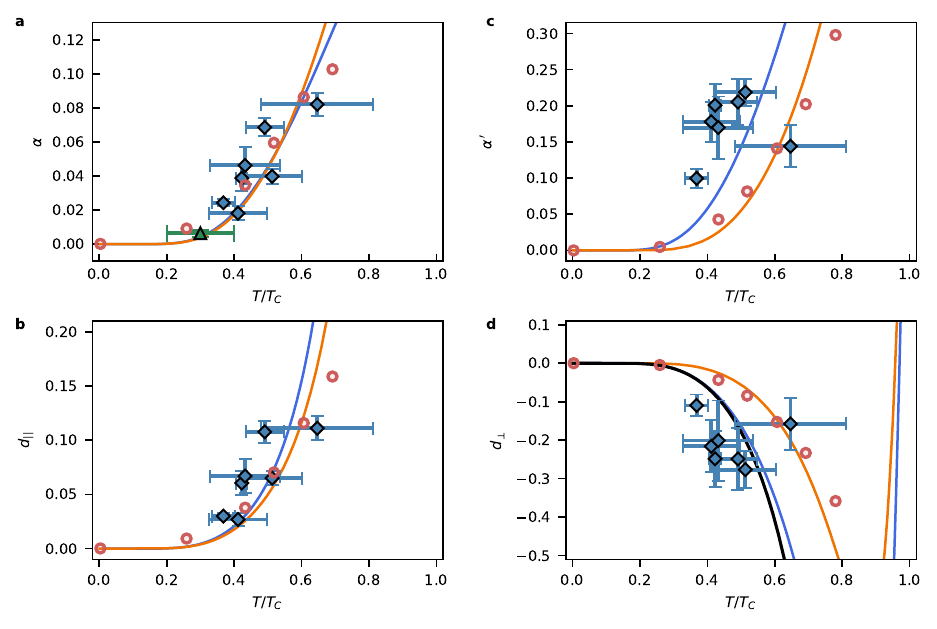}\vspace{-5 pt}
\caption{\textbf{Mutual friction coefficients and forces.} Temperature dependence of $\alpha$ \textbf{a)}, $\alpha'$ \textbf{b)}, $d_{||}$ \textbf{c)} and $d_{\perp}$ \textbf{d)}. Color code is the same of Fig.~\ref{fig:Fig3} c-d. The additional orange line corresponds to the estimation using Kopnin's model within the SLDA framework. The blue (orange) curve is obtained with $A=0.4$ ($A=1.5$) for the corresponding critical temperature $T_c=0.17\,T_F$ ($T_c=0.3\,T_F$). Black line represents the value expected from the Iordanskii force.
}
\label{fig:extFigds}
\end{figure*}

Assuming that the normal component is at rest, $\vec{v}_n=0$, and considering that the central vortex is ideally pinned at the origin, the motion of the satellite vortex is described by:
\begin{align}
\frac{d \vec{r}}{dt} &= (1-\alpha') \vec{v}^0_s - \alpha \sigma\hat z \times \vec{v}^0_s,\\
\vec{v}^0_s &= -\frac{\kappa_c}{2\pi} \hat z \times \left(\frac{\vec r}{|r|^2} + \frac{\vec r-\frac{R^2}{|r|^2} \vec r}{|\vec r-\frac{R^2}{|r|^2} \vec r|^2}
    \right),
\end{align}
where $\vec{v}^0_s$ is the superfluid velocity generated by the central pinned vortex and by the imaginary vortex. We note that the assumption $\vec{v}_n=0$ can be relaxed, at the expense of providing the solution of the coupled Navier-Stokes equations, or Boltzmann equation, for the normal component, thus obtaining $\vec{v}_n$. More complex approaches also simulating the normal component such as \cite{tang2023imaging, galantucci2020new, Jackson2009} may provide benchmarks on the effects of the back-reaction of the normal component to the vortex dynamics.

This set of equations can be solved analytically in the case $\alpha \neq 0$, yielding:
\begin{align}
 r(t) &= R \sqrt{\frac{1}{2}-\frac{1}{2}W\left(\left[1-2\left(\frac{r_0}{R}\right)^2\right]e^{1-2\left(\frac{r_0}{R}\right)^2+\frac{8\alpha\kappa}{R^2}t}\right)},\\
 \theta(t) &= \theta_0 + \frac{1-\alpha'}{\alpha \sigma} \log \left(\frac{r(t)}{r_0}\right),
\end{align}
where $W(x)$ is the Lambert W-Function, which is the inverse function of $f(W) = We^W$, and $(r_0,\theta_0)$ are the initial coordinates at $t=0$. Letting $\alpha = 0$ leads to the solution:
\begin{align}
r(t) &= r_0,\\
\theta(t) &= \theta_0-\sigma(1-\alpha')\kappa\left(\frac{2r_0^2-R^2}{r_0^4-r_0^2R^2}\right) t,
\end{align}
with a constant angular velocity as a function of time. The choice of $\sigma=\pm 1$ simply sets the direction of motion of the satellite vortex, clockwise or anticlockwise.

\subsection*{Analytical model of mutual friction for Fermi superfluids}
The values of the parameters $D$ and $D'$ can be theoretically predicted from the microscopic theory of the superfluid system and of the scattering processes. This, together with Eq.~\eqref{eq:alpha}-\eqref{eq:d_s}, set the vortex motion as a macroscopic probe for microscopic processes, thus providing a benchmark for microscopic theories. In Fig.~\ref{fig:Fig3}, we compare the obtained values of $\alpha$ and $\alpha'$ with the ones predicted by Kopnin and Volovik \cite{kopnin2002vortex,kopnin1995spectral} considering the effect of localized quasiparticles in the mutual friction. The mutual friction coefficients predicted by this model have the following temperature dependence:

\begin{align}\label{eq:Kopnin}
d_{||}  &= \frac{\rho}{\rho_s} \frac{\omega_0\tau}{1+\omega_0^2\tau^2} \tanh\left(\frac{\Delta}{2k_BT}\right),
\\
d_{\perp}  &= 1 - \frac{\rho}{\rho_s} \frac{\omega_0^2\tau^2}{1+\omega_0^2\tau^2} \tanh\left(\frac{\Delta}{2k_BT}\right),
\label{eq:d_perp_kopin}
\end{align}
where we introduced the superfluid fraction $\rho_s/\rho$. 
We estimate their values considering $\hbar \omega_0 = |\Delta|^2/E_F$, and $\tau = \frac{1}{A}\frac{\rho}{\rho_n}\tau_n$, with A being a phenomenological parameter of the order of 1, and $\tau_n$ is the Fermi-liquid relaxation rate in the normal state at the critical temperature, that we estimate as $\tau_n = E_F/T_c^2$ \cite{kopnin2002vortex}.
Let us remark that Eq.~\eqref{eq:d_perp_kopin}, already includes the presence of the Iordanskii force \cite{kopnin1991mutual} written as $d_{\perp} = -\rho_n/\rho_s = -(1-\rho_s)/\rho_s$. In Extended~Data~Fig.~\ref{fig:extFigds}, this term is explicitly shown in comparison with the rest of the curves. Substituting Eqs.~\eqref{eq:Kopnin}-\eqref{eq:d_perp_kopin} into Eqs.~\eqref{eq:alpha}-\eqref{eq:alphaprime}, without any approximations, we get:
\begin{align}
\alpha  &= \frac{1}{\omega_0\tau} \frac{\rho_s}{\rho}\frac{1}{\tanh\frac{|\Delta|}{2k_BT}},\\
\alpha' &= 1-  \frac{\rho_s}{\rho}\frac{1}{\tanh\frac{|\Delta|}{2k_BT}}.
\end{align}
Equivalently, 
\begin{align}
\omega_0\tau &= A\ \frac{\tilde{\Delta}^2}{1-\rho_s/\rho} ,\\
\alpha  &= \frac{1}{A} \frac{1}{\tilde{\Delta}^2} \frac{\rho_s}{\rho}\frac{1-\frac{\rho_s}{\rho}}{\tanh\frac{\tilde{\Delta}}{2\tilde{T}}},\\
\alpha' &= 1-  \frac{\rho_s}{\rho}\frac{1}{\tanh\frac{\tilde{\Delta}}{2\tilde{T}}},
\end{align}
where $\tilde{\Delta} = |\Delta|/k_BT_c$, $\tilde{T} = T/T_c$. Notice that depending on the value of $T_c/T_F$, the behavior of $\alpha$ and $\alpha'$ is expected to change.

The dissipative coefficient $\alpha$ depends on the parameter $A$, which can therefore be adjusted to match the experimental data. Moreover, there are no ab initio calculations that may give its exact value. The continuous blue line in Fig.~\ref{fig:Fig3}-\ref{fig:Fig4} of the main text is obtained with the value of $A=0.4$. Surprisingly, the non-dissipative coefficient $\alpha'$ is completely locked by the temperature dependence of the superfluid fraction, $\rho_s/\rho$, and the gap, $\Delta$. Therefore, there are no fitting parameters for $\alpha'$. Moreover, the ratio $(1-\alpha')/\alpha$ can also be calculated without approximations:
\begin{align}
\frac{1-\alpha'}{\alpha}  &= \frac{1-d_{\perp}}{d_{||}} = \omega_0\tau.
\end{align}

The relevance of the mutual friction coefficients is typically assigned according to their relevant scalings. For instance in Bose superfluids, near $T=0$, it is considered that $\alpha'$ scales as $\alpha^2$. Hence, for small values of $\alpha$, the effect of the non-dissipative coefficient is considered negligible. This is not necessarily the case for Fermi superfluids. The relative dependence between $\alpha$ and $\alpha'$ is given by:
\begin{equation}
\alpha' = 1 - \omega_0\tau\alpha.
\end{equation}
Since Fermi superfluids typically operate in the intermediate to super-clean regimes, where $\omega_0\tau \gtrsim 1$, even when $\alpha$ is small, $\alpha'$ could be of the order of unity. Indeed, we find $\alpha'\sim\alpha^{1.3(1)}$ for the theoretical models shown in Extended Data Fig. \ref{fig:extFigds}, for $T>0.3T_c$.

\subsection{SLDA simulations}

\begin{figure}[t!]
\centering
\vspace{0 pt}
\includegraphics{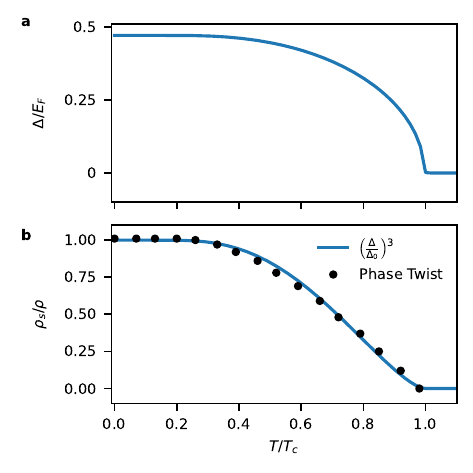}\vspace{-5 pt}
\caption{\textbf{Thermodynamic properties from the SLDA.}
\textbf{a)} Gap $\Delta$ as a function of temperature
\textbf{b)} Estimated superfluid fraction obtained from the phase twist method, and phenomenological behavior expressed in terms of the reduced gap $\Delta/\Delta_0$.
}
\label{fig:extFig4}
\end{figure}

The density-functional theory we apply here is formally equivalent to the mean-field Bogoliubov–de Gennes equations~\cite{Bulgac2007}. For static problems, they have the form ($\hbar=m=k_B=1$)
\begin{equation}\label{eq:sldaGB}
    \mathcal{H}(\rho,\nu) \binom{u_n(\vec{r})}{v_n(\vec{r})} = E_n\binom{u_n(\vec{r})}{v_n(\vec{r})},
\end{equation}
and describe the Bogoliubov amplitudes $(u_n(\vec{r}),v_n(\vec{r}))^T$ that in turn define densities: normal $\rho$, anomalous $\nu$ and current $\vec{j}$ 
\begin{align}
    \rho(\vec{r})   &= 2\sum_{E_n>0} |v_n(\vec{r})|^2 f^-_n + u_n(\vec{r})|^2 f^+_n, \label{eq:rho}\\
    \nu(\vec{r}) &= \sum_{E_n>0} (f^-_n -f^+_n) u_n(\vec{r})v^*_n(\vec{r}),\\
    \vec{j}(\vec{r}) &= 2\sum_{E_n>0}\left(  \textrm{Im}[v_{n}(\vec{r})\nabla v_{n}^*(\vec{r})] f^-_n\right.\nonumber \\
& \quad\quad\quad\quad - \left. \textrm{Im}[u_{n}(\vec{r})\nabla u_{n}^*(\vec{r})] f^+_n \right).\label{eq:current}
\end{align}
The temperature effects are modeled by introducing the Fermi-Dirac distribution, noted as $ f^{\pm}_n = (1+\exp(\pm E_n/T))^{-1}$, when computing densities from the Bogoliubov amplitudes, and the framework becomes formally equivalent to the finite-temperature HFB method~\cite{Goodman1981}. The time-dependent equations are obtained by changing $E_n\rightarrow i\frac{\partial}{\partial t}$ and allowing all functions to be time-dependent as well. The Hamiltonian has the generic form
\begin{equation}
\mathcal{H} =
    \begin{pmatrix}
     -\frac{1}{2}\nabla^2+U(\vec{r})-\mu &  \Delta(\vec{r}) \\
      \Delta^*(\vec{r})& \frac{1}{2}\nabla^2-U(\vec{r})+\mu
    \end{pmatrix},
\end{equation}
where mean and pairing fields are computed as appropriate functional derivatives of the SLDA functional~\cite{Bulgac2007}. Their explicit forms are:
\begin{align}
U &= \frac{\hat\beta (3\pi^2\rho)^{2/3}}{2} - \frac{|\Delta|^2}{3\hat\gamma \rho^{2/3}}+ V_{\textrm{ext}},\\
\Delta &= -\frac{\hat\gamma}{\rho^{1/3}}\nu.
\end{align}
The functional parameters ($\hat \beta$ and $\hat \gamma$) are adjusted to ensure the reproduction of the Bertsch parameter $\xi\approx0.4$ and the pairing gap at zero temperature $\Delta/\varepsilon_F\approx0.5$ of the uniform unitary Fermi gas, consistent with the quantum Monte Carlo results~\cite{Bulgac2012}. One of the drawbacks of such a defined framework is an overestimation of the critical temperature of the superfluid-to-normal phase transition, which reads $T_c\approx0.305\,T_F$. The temperature dependence of the pairing gap, for uniform UFG, is shown in the Extended~Data~Fig.~\ref{fig:extFig4}a. 

To solve static and time-dependent equations, we have used \verb|W-SLDA Toolkit|~\cite{WSLDAToolkit}, and calculations were executed on the LUMI supercomputer (Kajaani, Finland). 

We estimated the superfluid fraction from the SLDA simulations using the method based on the response of the system to a phase twist \cite{richaud2024dynamical} and relying on a concept of Landau's two-fluid model. We assume that the density~(\ref{eq:rho}) and current~(\ref{eq:current}) can be decomposed into normal ($n$) and superfluid ($s$) parts:
\begin{align}
    \rho &= \rho_s + \rho_n,\\
    \vec{j} &= \vec{j}_{s} + \vec{j}_{n} = \rho_s\vec{v}_{s} +  \rho_n\vec{v}_{n},
\end{align}
where the superfluid velocity $\vec{v}_{s}$ is related to the gradient of the phase of the order parameter $\vec{v}_s(\vec{r},t) =\frac{1}{2}\nabla \phi(\vec{r},t)$ (we used mass of Cooper pair $M=2m=2$ in our units). We consider that the normal component is at rest with the system boundaries, $\vec{v}_{n} = 0$. Therefore, the total density current $\vec{j}$ reduces to $\rho_s\vec{v}_{s}$. We can estimate the superfluid fraction in the bulk by computing the ratio:
\begin{equation}
    \frac{\rho_s}{\rho} = \frac{\frac{|\vec{j}|}{|\vec{v}_s|}}{\rho}= \frac{2|\vec{j}|}{\rho|\nabla \phi|}.
    \label{eq:fs}
\end{equation} 
In the Extended~Data~Fig.~\ref{fig:extFig4}b, we show the extracted superfluid fraction as a temperature function by considering a uniform system with imposed superfluid flow in one direction. The obtained values turn out to be well approximated by the phenomenological formula $\rho_s/\rho = \left(\Delta(T)/\Delta(T=0)\right)^{3}$. 

\section{Data availability}
The data that support the figures within this paper are available from the corresponding author upon reasonable request.

\section{Author contributions} 
N.G., D.H.R, and G.R. conceived the study.
N.G., D.H.R, and C.D. performed the experiments. 
N.G., D.H.R, and C.D. analyzed the experimental data. 
D.H.R, G.W, P.M. carried out the SLDA numerical calculations. 
P.P. and M.P. carried out the self-consistent $t$-matrix calculations. 
All authors contributed to the interpretation of the results and to the writing of the manuscript.

\section{Competing interests} 
The authors declare no competing interests.

%


\renewcommand{\figurename}{Supplementary Figure}
\renewcommand{\thefigure}{S.\arabic{figure}}
\setcounter{figure}{0}
\renewcommand{\theequation}{S.\arabic{equation}}
\setcounter{equation}{0}
\renewcommand{\thesection}{S.\arabic{section}}
\setcounter{section}{0}
\renewcommand{\thetable}{S.\arabic{table}}
\setcounter{table}{0}

\setlength{\tabcolsep}{18pt}

\onecolumngrid
\setcounter{equation}{0}
\setcounter{figure}{0}
\setcounter{table}{0}

\clearpage

\begin{center}
\textbf{\large Supplementary Information}
\end{center}

\subsection{Deterministic vortex dipole creation}
To deterministically create a vortex dipole we implement the chopstick technique for vortex creation proposed by \cite{Samson_DeterministicCreation}, and already implemented in our previous work \cite{kwon2021sound}. Both the in-plane circular box potential and the chopsticks are implemented using a single DMD. We dynamically control the potential projecting with the DMD a sequence of images  with controlled timing. In contrast with other methods for the creation of vortices, the chopstick method has the advantage of controlling each vortex individually using two focused repulsive obstacles as effective pinning centers for the vortex-antivortex pair. The pristine control offered by the chopstick method, allow us to fine tune the vortex dipole initial size, position and orientation. Contrarily to \cite{kwon2021sound}, we now accommodate the antivortex at the center of the disk and keep its pinning potential on for all the vortex dynamic, which we trigger by releasing the pinning potential of the off-centered vortex.

After moving the pinning potentials to their final locations, both vortices remain pinned for $\SI{30}{ms}$ to allow the dissipation of any spurious excitations. The removal time of $\SI{1}{ms}$ is chosen to be sufficiently slow to minimize sound excitations that could affect the initial vortex motion, but fast enough to prevent uncontrolled vortex depinning, thus maintaining a high experimental shot-to-shot reproducibility.

\subsection{Vortex imaging}
To image the vortices we acquire a time-of-flight (TOF) image of the superfluid density. In particular, we abruptly switch off the vertical confinement and at the same time, we start to ramp down the DMD potential, removing it completely in $\SI{1}{ms}$. Then, we let the system evolve further for $\mathrm{0.7} - \SI{1.2}{ms}$ of TOF and then acquire the absorption image. Simultaneously, we linearly ramp the magnetic field to $\SI{630}{G}$ in $\SI{2.7}{ms}$ to map the system in a BEC superfluid, where vortices appear as clear holes in the TOF expansion. The ramp of the magnetic field ends when the image is acquired. This TOF protocol allows for maximization of the vortex visibility while keeping the size of the vortex core small enough to precisely locate their position with sufficient accuracy. To estimate the position of each vortex, we perform a Gaussian fit around the vortex density depletion.

\begin{figure*}[ht!]
\centering
\vspace{0 pt}
\includegraphics{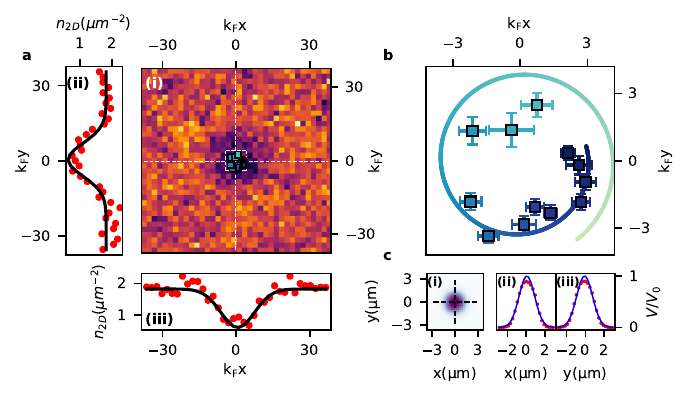}\vspace{-5 pt}
\caption{\textbf{Pinning of the central vortex.}
(\textbf{a}). (i) Zoom of the region around the central vortex in Fig.2a (ii). (ii-iii) Vertical and horizontal density cut of the central region of the vortex (white dashed lines in (i)). The position of the vortex is obtained from a 2-dimensional Gaussian fit (black continuous lines). Square symbols in (i) represent the observed average position on the central vortex as a function of time. (\textbf{b}) Time evolution of the observed position of the central vortex. The plot shows the region delimited by the white dashed square in panel \textbf{a}. The continuous line displays the expected position from the TOF procedure. Colors encode evolution times, with the same color scale as in Fig.2 of the Main. (\textbf{c}) Profile of the optical potential used to keep the central vortex pinned. Black dashed lines in (i) mark horizontal and vertical cuts across the central position, shown in (ii) and (iii) respectively. Red dots are measurements, while the blue continuous line is a Gaussian fit.
}
\label{fig:extFig2}
\end{figure*}

\subsection{Keeping the central vortex pinned}

The initial position of the vortex dynamics is fixed by the position of the chopsticks before turning them off. We study the stability of the initial configuration by observing the position of the vortices as a function of time keeping the two optical potentials on in a typical experimental configuration. We observe that the position of the two vortices is fixed for time $\sim \SI{1}{s} \gg \SI{30}{ms}$ we wait before trigger the dynamics, demonstrating our capability in pin the vortex position in the initial configuration.

We also monitor the pinned vortex position during the orbiting dynamic of the mobile vortex, as it is shown in Supplementary~Fig.~\ref{fig:extFig2}, the inner vortex is observed in different positions at different evolution times with displacements on the same size of the vortex hole in the TOF images. Despite the short time of the TOF protocol, this behavior can be associated with the imaging procedure, during which the pinning potential is removed. Consequently, the vortex-antivortex configuration moves as a free dipole \cite{kwon2021sound} for $\SI{2.7}{ms}$, yielding to a displacement of the inner vortex that we estimate in the following way. After an evolution time $t$, the two vortices are in coordinates (0,0) and ($r(t) \cos{\theta(t)}$,$r(t) \sin{\theta(t)}$). The expected displacement during the TOF is calculated from the propagation without boundaries but subject to the same mutual friction coefficients as the measured \textit{in situ} ones. Following the Eq.~(9), the position of the central vortex ($x_{in}$,$y_{in}$) after a propagation as a free dipole for a time $t_{TOF}$ is given by:
\begin{equation}
\label{x_innervortex}
x_{in}(t,t_{TOF}) = y_0(t,t_{TOF}) \cos{(\theta(t))} + x_0(t,t_{TOF}) \sin{(\theta(t))}
\end{equation}
\begin{equation}
\label{y_innervortex}
y_{in}(t,t_{TOF}) = y_0(t,t_{TOF}) \sin{(\theta(t))} - x_0(t,t_{TOF}) \cos{(\theta(t))},
\end{equation}
with 
\begin{equation}
\label{x0_innervortex}
x_0(t,t_{TOF}) = \frac{1-\alpha ' }{\alpha} y_0(t,t_{TOF})
\end{equation}
\begin{equation}
\label{y0_innervortex}
y_0(t,t_{TOF}) = r(t) -\sqrt{r(t)^2 - \alpha \frac{\hbar}{2m}t_{TOF}}. 
\end{equation}
Supp.~Fig.~\ref{fig:extFig2}b shows the observed position of the inner vortex for the experimental realization in Fig.~2 of the main text (squared points). The colored solid line represents the estimated position of the inner vortex after a $t_{TOF} = \SI{2.7}{ms}$. Different colors encode the time evolution $t$ with the same color map as in Fig.~2b of the main text. The agreement between the observed position and the one estimated with Eq.~\eqref{x_innervortex}-\eqref{y_innervortex} suggests that the motion of the inner vortex can be fully associated with the TOF procedure. Moreover, it proves our capability to track the vortex position at sizes smaller than the vortex hole size after the TOF procedure, with 1/e radius $\sigma = \SI{2.0 \pm 0.2}{\mu m}$. During the free dipole evolution, the angular relative position between the two vortices is preserved, while the dipole shrinks due the effect of $\alpha$. Considering Eq.~\eqref{x_innervortex}-\eqref{y_innervortex}, the radial distance decreases by a factor $2 y_0(t,t_{TOF})$, that we can estimate to be $\mathrm{0.04}-\SI{0.4}{\mu m}$, much smaller than the typical distances observed in our experiment $r(t) \sim \SI{10}{\mu m}$. As a consequence the off center vortex position is mapped to the relative position between the two vortex, that we take as main observable for all data presented in the main text. Supp. Fig. ~\ref{fig:extFig2}c(i) shows the profile of the pinning potential used to fix the position of the central vortex. (ii)-(iii) show the horizontal and vertical cut corresponding to the black dashed lines in (i). The potential profile is fitted by a Gaussian function obtaining $V_0 =\SI{22 \pm 3}{kHz}$ and $1/e^2$ radius of $\sigma = \SI{1.6\pm 0.2}{\mu m}$.

\subsection{Fitting procedure to extract $\alpha$ and $\alpha'$}
\begin{figure}[ht!]
\centering
\vspace{0 pt}
\includegraphics{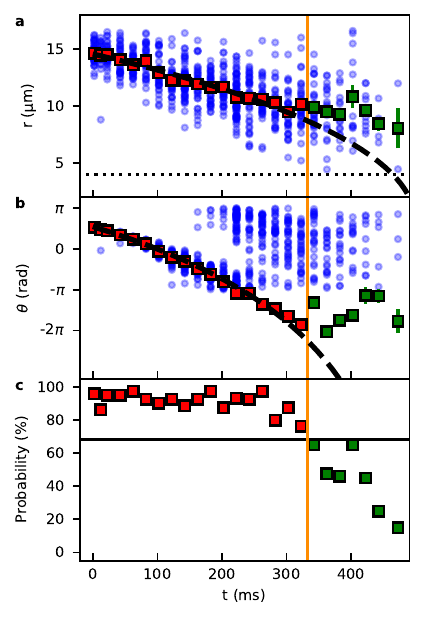}\vspace{-5 pt}
\caption{\textbf{Fitting of the vortex trajectories.} \textbf{a-b} radial and azimuthal trajectory as a function of time. Blue circles represent the observed values for different experimental realizations. Green and red squares are the average values. Black dashed lines are the fit using DPVM. \textbf{c} Percentage of experimental runs considered in the fitting procedure. Black continuous line mark our cutoff value ($0.7 \times \mathrm{max(Probability)}$) to consider the point in the fit. Red and green points represent the points considered and excluded from the fitting procedure, respectively. The orange vertical line sets the separation between the two groups of points.
}
\label{fig:extFig3}
\end{figure}

To extract the value of $\alpha$ and $\alpha '$ we fit the radial and angular time evolution of the relative position between the two vortices as a function of time. Supp.~Fig.~\ref{fig:extFig3} shows an example of the typical procedure. The value and the error of the radial position in time is calculated from the average and standard deviation of the mean of the values of different experimental realizations. The value of $\theta(t)$ is obtained from the circular mean of the values obtained over different repetitions, and subsequently unwrapped out of the range ($-\pi$,$\pi$), while the associated error is calculated from the circular standard deviation divided by the squared root of the number of repetitions. Supp.~Fig.~\ref{fig:extFig3}(\textbf{c}) shows the percentage of experimental shots considered for the fitting procedure. In particular, we exclude the cases in which the vortices are absent, depinned, or when the determination of the position via a gaussian fit of the density depletions are not reliable because of noise in the image. Such a probability decreases as time proceed. This behavior could bring to an overestimation of the radial vortex position, as annihilation and depinning processes are more favorable to happen at short distance, removing low values of the radial position in the statistical ensemble from which the average is estimated. To avoid effect on the estimation of the mutual friction coefficients, we restrict our analysis for evolution times with dipole probability higher than the $70\%$ of the maximum probability at different times, which we found to be at least $95\%$. This threshold is marked by the continuous black horizontal line in (\textbf{c}). The vertical orange line separates the points considered in the fitting procedure (red squares) from the one neglected (green squares). As it is shown in (\textbf{a}) the green points represent an overestimation of the expected radial evolution of the fit (black dashed line). Considering these points would lead to an underestimation of the value of $\alpha$. Moreover, we do not observe pinned dipoles at distances $<\SI{4}{\mu m} \sim 10 k_F^{-1}$, marked by the horizontal dotted black line.

\subsection{Quasi-2D vortex dynamics}
The vertical confinement provided by the $\mathrm{TEM_{01}}$ laser beam is such that $\mu/\hbar \omega_z \approx 7$, making the system collisionally three-dimensional. However, vortex dynamics behave as a quasi-two-dimensional system since only a few Kelvin modes can be populated. The standard Kelvin dispersion \cite{Soninbook2015} is:
\begin{equation*}
    \omega(k) = -\frac{\hbar k^2}{4m} \log\left(\xi k\right),
\end{equation*}
where $\xi$ is the healing length. Due to geometric restrictions \cite{Soninbook2015}, only the modes with a wave vector $k \xi_v \lesssim 1$ can be effectively populated in the superfluid. Under our experimental condition, this translates into the fact that only the lowest four Kelvin modes with $k_n=(\pi n)/2R_z$ can be thermally populated, where $R_z=\sqrt{2\mu/(2m\omega_z^2)}$ is the Thomas-Fermi radius in the $\hat{z}$ direction. Given the small number of possibly populated Kelvin modes, we can assume a 2D dynamics of the vortex motion. Moreover, in the vortex images we do not see any elliptical distortion of the circular vortex profile.

\subsection{Critical temperature  of unitary fermions in a hybrid trap}

We estimate the critical temperature of the unitary gas in the hybrid trap of the experiment by means of a fully self-consistent $t$-matrix approach \cite{Haussmann1993,Haussmann1994,Pini2019}.

To this end, we consider a spin-balanced Fermi gas at temperature $T$ made of $N=N_{\uparrow}+N_{\downarrow}$ atoms with mass $m$ trapped in a potential that, in cylindrical coordinates, is given by
\begin{equation}
\label{eq:pancake_potential}
V(r,z) = \frac{1}{2} m \omega_z^{2} z^{2} +V_0 \, \Theta(r-R),
\end{equation}
where $\omega_z$ is the angular frequency of the harmonic potential along the $z$ direction and $V_0 \to +\infty$ is a hard-wall potential of radius $R$. Within this potential, the Fermi energy
reads
\begin{equation}
    E_F=\left[\frac{2}{m} \left( \frac{\hbar}{R} \right)^2 \hbar \omega_z N \right]^{1/2}.
\end{equation}
To describe the system confined by the potential (\ref{eq:pancake_potential}), we adopt a local density approximation, treating the system as locally homogeneous. The chemical potential $\mu$ is replaced with a position-dependent chemical potential $\mu(r,z)=\mu_0-V(r,z)$, thus obtaining a local fermionic Green's function $G(\mathbf{k},\omega_n;\mu(r,z))$.
From the local Green's function, one obtains the number density profile
\begin{equation}
\label{eq:density_equation_trap}
\rho(r,z)=2 \int \frac{d\mathbf{k}}{(2\pi)^3} \frac{1}{\beta} \sum_n G(\mathbf{k},\omega_n; \mu(r,z)),
\end{equation}
where $\beta=1/(k_B T)$ and $\omega_n=\pi T (2n+1)$ (with $n$ integer) is a fermionic Matsubara frequency. Note that here the $r$-dependence is trivial: $\rho(r,z)=\Theta(R-r)\rho(z)$. 
The chemical potential $\mu_0=\mu(r=0,z=0)$ can then be determined  by inverting the number equation
\begin{equation}
\label{eq:number_equation_trap}
N= 2\pi R^2  \int_{0}^{\infty} \! dz  \, \lambda(z),
\end{equation}
where $\rho(z)$ is calculated on an appropriate numerical grid. For a given local chemical potential $\mu$, the single-particle Green's function $G(\mathbf{k},\omega_{n})$ is calculated within the self-consistent $t$-matrix approach (see \cite{Haussmann1993,Haussmann1994,Pini2019} for details on its implementation, and \cite{Haussmann2008,Pini2020,Pini2021} for its applications to trapped systems). It reads:
\begin{equation}
G(\mathbf{k},\omega_{n}) = \left[ G_{0}(\mathbf{k},\omega_{n})^{-1} - \Sigma(\mathbf{k},\omega_{n}) \right]^{-1},
\label{eq:G}
\end{equation}
where $G_{0}(\mathbf{k},\omega_{n})^{-1} = i \omega_{n} - \hbar\mathbf{k}^{2}/(2m) +\mu/\hbar$, while 
\begin{equation}
\Sigma(\mathbf{k},\omega_{n}) = - \int \! \frac{d{\mathbf q}}{(2\pi)^{3}} \frac{1}{\beta} \sum_{\nu} \Gamma(\mathbf{q},\Omega_{\nu}) \,G(\mathbf{q}-\mathbf{k},\Omega_{\nu}-\omega_{n}) 
\label{eq:t-matrix-self-energy}
\end{equation}
is the self-energy, where $\Omega_\nu=2\pi T \nu$ (with $\nu$ integer) is a bosonic Matsubara frequency. The particle-particle propagator $\Gamma(\mathbf{q},\Omega_{\nu})$ in Eq.~(\ref{eq:t-matrix-self-energy}) is given by 
\begin{equation}
\label{eq:Gamma}
\Gamma(\mathbf{q},\Omega_{\nu})= - \bigg[\frac{m}{4\pi\hbar^2 a} + R_{pp}(\mathbf{q},\Omega_{\nu})\bigg]^{-1},
\end{equation}
where $a$ is the s-wave scattering length, while the renormalized particle-particle bubble $R_{pp}(\mathbf{q},\Omega_{\nu})$ is given by
\begin{align}
\label{eq:Rpp}
R_{pp}(\mathbf{q},\Omega_{\nu})&= \int\!\!\frac{d\mathbf{k}}{(2\pi)^3} \bigg[ \frac{1}{\beta} \sum_{\omega_n} G(\mathbf{q}+\mathbf{k},\Omega_{\nu}+\omega_{n})   \nonumber\\
&\times G(-\mathbf{k},-\omega_n)- \frac{m}{\hbar^2\mathbf{k}^2}\bigg].    
\end{align}
At unitarity $1/a \to 0$ and $\Gamma(\mathbf{q},\Omega_{\nu})= - 1/R_{pp}(\mathbf{q},\Omega_{\nu})$. The critical temperature is determined by requiring that the Thouless criterion \cite{Thouless1960} is satisfied at the center of the trap, which corresponds to the condition
\begin{equation}
    \label{eq:Thouless}
    [\Gamma(\mathbf{q}=0,\Omega_\nu=0;\mu=\mu_0)]^{-1}=0.
\end{equation}

Equations (\ref{eq:G})-(\ref{eq:Rpp}) for the single-particle Green's function, together with the Thouless criterion (\ref{eq:Thouless}) and the number equation (\ref{eq:number_equation_trap}) are solved self-consistently. The numerical calculation of the expressions (\ref{eq:G})-(\ref{eq:Rpp}) is implemented taking advantage of the procedures detailed in~\cite{Pini2019}. In this way, for the trapping potential (\ref{eq:pancake_potential}), we obtain a critical temperature $T_c/T_F=0.180$ at unitarity.

\subsection{SLDA simulations}

To simulate the vortex in the SLDA simulations, we impose an external potential $V_{\textrm{ext}}(\mathbf{r})= V_{\textrm{box}}(\mathbf{r})+ V_{\textrm{pin}}(\mathbf{r})$. The first entry represents the potential of the confining box in the plane $xy$, which is zero inside the radius cylinder $r=\sqrt{x^2+y^2}<R_{\textrm{box}}$ and it increases to value $10E_F$ otherwise. The second one represents the pinning potential $V_{\textrm{pin}}$, and is given as a set of two Gaussian potentials of width $\sigma = 2k_F^{-1}$, and height $V_{\textrm{pin}} = E_F$ located at the origin and at a distance $d_0$ varied depending of the system size. The chemical potential $\mu$ is adjusted in such a way as to obtain the desired value of the bulk density $\rho_0$, which defines the Fermi wave vector $k_F=(3\pi^2 \rho_0)^{1/3}$ and the Fermi energy $E_F=k_F^2/2$. 

\begin{figure*}[t!]
\centering
\vspace{0 pt}
\includegraphics{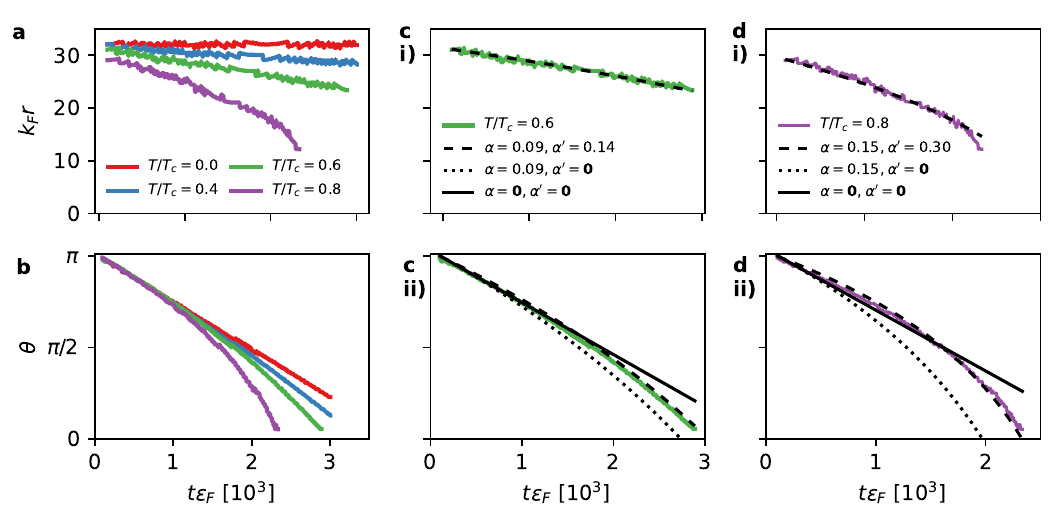}\vspace{-5 pt}
\caption{\textbf{Vortex trajectories obtained from the SLDA.}
\textbf{a)} Size and \textbf{b)} angular position of the vortex dipole as a function of time, for different temperatures. Comparison of vortex trajectories for \textbf{c)} $T/T_c=0.6$ and  \textbf{d)} $T/T_c=0.8$ with the DPVM assuming different values of $\alpha$ and $\alpha'$.
}
\label{fig:extFig5}
\end{figure*}
The equations of motion~(28) of the main text were solved in the coordinate space on a mesh of size $N_x\times N_y\times N_z$, with grid spacing $\Delta x k_F=1$. Since we do not include in our numerical simulations confining in the $z$ direction, we have assumed that the wave functions are plane waves in this direction, namely $u_n(\mathbf{r})=u_n(x,y)e^{ik_z z}$ and $v_n(\mathbf{r})=v_n(x,y)e^{ik_z z}$, where $k_z=0, \pm 1\frac{2\pi}{L_z},  \pm 2\frac{2\pi}{L_z}, \ldots, \pm \frac{\pi}{\Delta x}$ with $L_z=N_z\Delta x$. This assumption freezes out dynamics in the $z$ direction, like the bending of vortex lines, while keeping the thermodynamic properties of the system as for the 3D case. In the computation, we have used $N_z=16$. For the perpendicular directions, we have tested cases $N_x=N_y = 128$, $196$, and $256$, corresponding to three different values of $R_{\textrm{box}}k_F=57.6$, $88.2$, and $115.2$. Given the different box radii, we were able to simulate three different dipole sizes, $d_0 k_F=21$, $31$, and $55$, for the corresponding grid sizes.

The computation consists of two steps: i) solving the static problem to generate the initial configuration and ii) performing the temporal evolution of the system. The initial configuration representing the vortex dipole pinned by the external potential $V_{\textrm{pin}}$ is obtained by means of the phase imprinting technique. In the process of finding the self-consistent solution of Eq.~(28) of the main text, we impose the phase $\phi$ of the order parameter $\Delta(x,y) = |\Delta(x,y)|e^{i\phi(x,y)}$ to be 
\begin{equation}
    \phi(x,y) = \tan^{-1}\left(\frac{y}{x}\right)  - \tan^{-1}\left(\frac{y}{x-d_0}\right).
\end{equation}
The initial positions of the vortices, $(0,0)$ and $(d_0,0)$, coincide with the locations of the Gaussian potentials in $V_{\textrm{pin}}$. The obtained quasi-particle wave functions $\{u_n, v_n\}$ are next evolved in time with the Adams-Bashforth-Moulton integration scheme of $5^{\textrm{th}}$ order, with integration time step $\Delta t =0.0035 \, E_F^{-1}$. The pinning potential of the off-centered vortex is removed following a sigmoid function shape in time with characteristic time $50 \, E_F^{-1}$, after which the vortex stars to orbit. We have used \verb|W-SLDA Toolkit|~\cite{WSLDAToolkit}, and calculations were executed on the LUMI supercomputer (Kajaani, Finland). 

\begin{figure*}[t!]
\centering
\vspace{0 pt}
\includegraphics{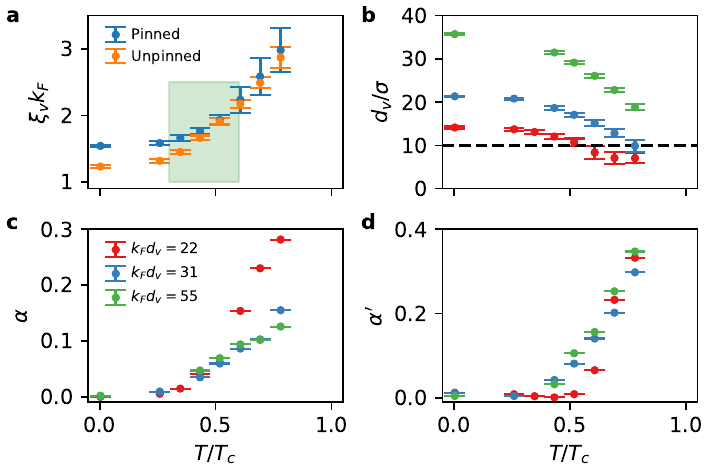}\vspace{-5 pt}
\caption{\textbf{Vortex and Dipole size effects on the SLDA simulations.} \textbf{a)} Core size of the pinned and unpinned vortex as a function of temperature. The shaded area shows the experimental region explored. \textbf{b)} Ratio of vortex dipole size $d_v$ to the vortex core size $\xi_v$. We test $k_Fd_v=22$, $31$, and $51$; the color code is in panel \textbf{c)}. For different dipole sizes, we measured \textbf{c)} $\alpha$,  and \textbf{d)} $\alpha'$ as a function of dipole size and temperature.
}
\label{fig:extFig6}
\end{figure*}

To track the position of the vortices, we analyze the phase of the order parameter. We calculate the rotation of the superfluid velocity $\textbf{v}_s$, and obtain the position of non-zero value, as shown in Supp. Fig. ~\ref{fig:extFig5} for different temperatures for a dipole $d_v=31 \, k_F^{-1}$. We also show the curves corresponding to the fit of the DPVM using different combinations of $\alpha$ and $\alpha'$.

From the SLDA simulations, we can get insight into the vortex's internal structure. In particular, the vortex core radius, $\xi_v$, changes with temperature, as shown in Supp. Fig. \ref{fig:extFig6}a. For comparison, we show the vortex core size pinned and unpinned by the Gaussian potential. Here, the vortex core size is obtained by fitting $\Delta(r)$ using a gaussian function $\exp(-r^2/2s_r^2)$.

\begin{figure}[ht!]
\centering
\vspace{0 pt}
\includegraphics[width=0.9\textwidth]{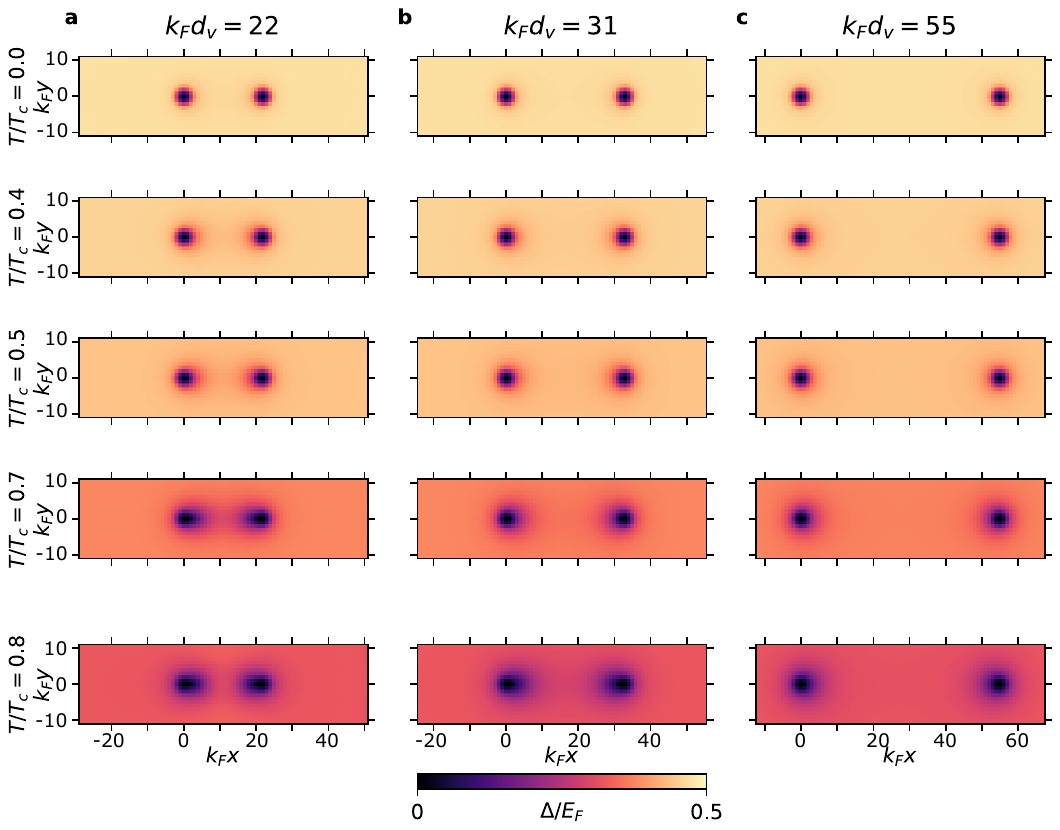}\vspace{-5 pt}
\caption{
\textbf{Order parameter $|\Delta|$ for different dipole sizes.} \textbf{a)} $k_Fd_v=22$, \textbf{b)} $k_Fd_v=31$, and \textbf{c)} $k_Fd_v=51$. Different temperatures are displayed vertically.
}
\label{fig:extFig7}
\end{figure}

For the latter scenario, the pinning potential widens the vortex core to the size of the Gaussian width. The fact that the vortex core changes with temperature also limits the range of applicability of the point vortex model to this system. In particular, when the vortices are separated by a distance $d_v\lesssim 10 \,\xi_v$, shown as a dashed line in Supp. Fig. \ref{fig:extFig6}b, the observed vortex dynamics are modified. 
In Supp. Fig. \ref{fig:extFig6}c-d, we show the fitted values of $\alpha$ and $\alpha'$ for all considered temperature and dipole size combinations. We observe that there is an increase of the coefficient $\alpha$ that deviates from the largest dipole probe when $d_v\lesssim 10 \,\xi_v$. We attribute such effect to an additional attraction caused by modification of the order parameter of one vortex by the close presence of the other; see Supp. Fig.~\ref{fig:extFig7}. This effect favors recombination of the vortices, increasing the energy dissipation, or equivalently the $\alpha$ parameter. We rely on larger dipole sizes to resolve this issue so each vortex can be considered to be immersed in a homogeneous background.

\begin{figure}[t!]
\centering
\vspace{0 pt}
\includegraphics{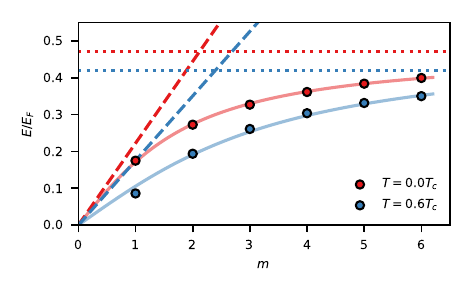}\vspace{-5 pt}
\caption{\textbf{Spectra of CdGM states in the Unitary Fermi gas}. Dispersion relation of CdGM states in the unitary Fermi gas at different temperatures obtained from SLDA. Dotted lines correspond to the gap $\Delta(T)$, dashed lines correspond to the approximation $|\Delta|^2/E_F$.
}
\label{fig:CdGM_spectra_Fig}
\end{figure}

\begin{figure}[ht!]
\centering
\vspace{0 pt}
\includegraphics[width=0.9\textwidth]{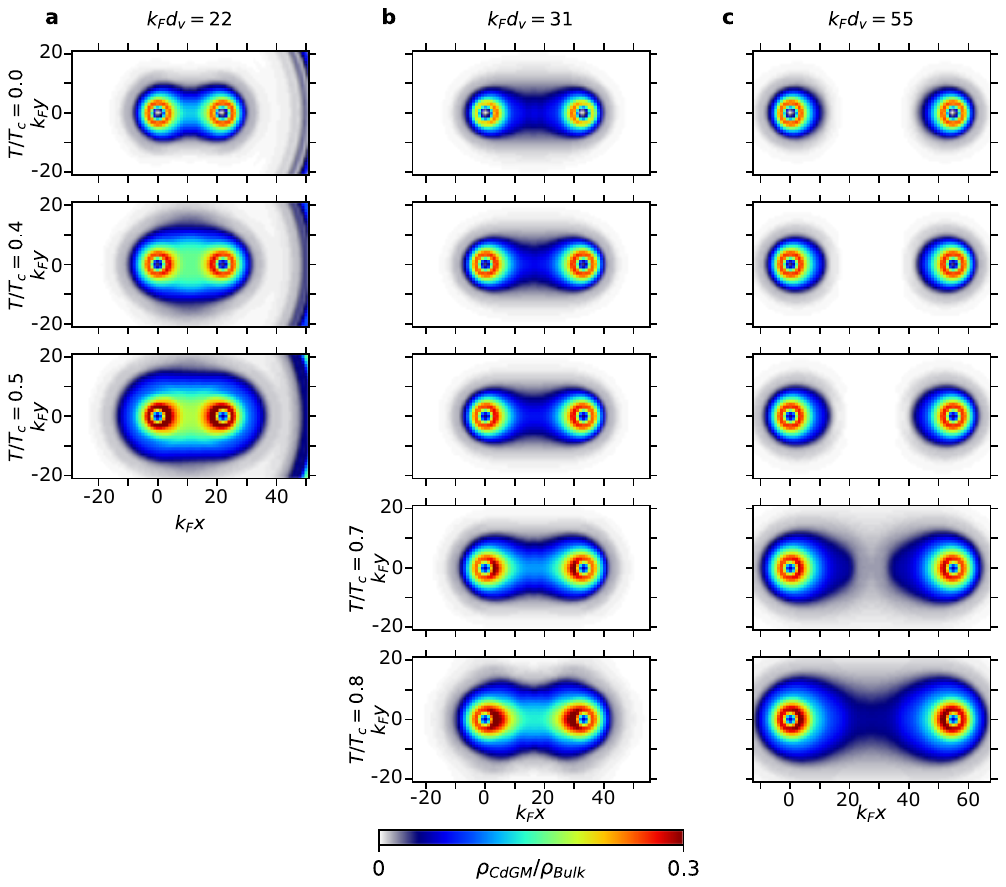}
\vspace{-5 pt}
\caption{
\textbf{Population of CdGM states $\rho_{\textrm{CdGM}}$ at time t=0}. At $t=0$, both vortices remain pinned by a Gaussian potential. We show the population of CdGM states for different dipole sizes \textbf{a)} $k_Fd_v=22$, \textbf{b)} $k_Fd_v=31$, and \textbf{c)} $k_Fd_v=51$. Different temperatures are displayed vertically.
}
\label{fig:extFig8}
\end{figure}

\begin{figure}[ht!]
\centering
\vspace{0 pt}
\includegraphics[width=0.9\textwidth]{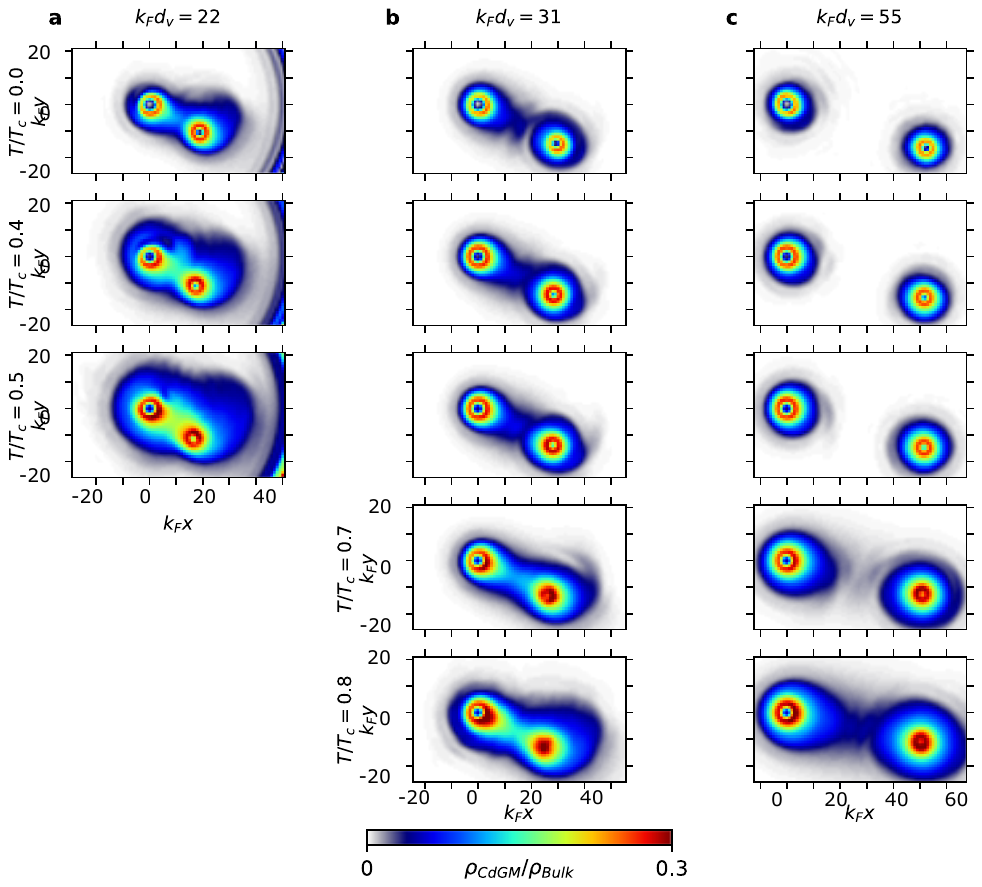}
\vspace{-5 pt}
\caption{
\textbf{Population of CdGM states $\rho_{\textrm{CdGM}}$ of unpinned vortex}. We show the population of CdGM states for different dipole sizes at different evolution times \textbf{a)} $k_Fd_v=22$ at $tE_F=300$, \textbf{b)} $k_Fd_v=31$ at $tE_F=600$, and \textbf{c)} $k_Fd_v=51$ at $tE_F=1200$. The time was chosen to match the angular position of the mobile vortex. Different temperatures are displayed vertically.}
\label{fig:extFig9}
\end{figure}

It is also instructive to analyze the spatial distribution of CdGM states. We define the density of the anomalous branch of the CdGM states in the context of the SLDA as
\begin{align}
    \rho_{\textrm{CdGM}}(\mathbf{r})   &= 2\sum_{0<E_n<0.8\Delta} |u_n(\mathbf{r})|^2 f^+_n + |v_n(\mathbf{r})|^2 f^-_n.
\end{align}
We impose an energy cutoff of $0.8\Delta$ to secure states with energies lower than the gap and get rid of states belonging to the first branch of CdGM states, which are hard to separate from the continuum states~\cite{barresi2023dissipative}. In the Supp. Figs. ~\ref{fig:extFig8} and \ref{fig:extFig9}, we show examples of such density distributions at $t=0$, when both vortices are pinned, and at $t>0$ when the off-centered vortex has already moved, while in Supp. Fig. \ref{fig:CdGM_spectra_Fig} we show the spectra of CdGM states for different temperatures. The spatial distribution of such states reveals that when $d_v\sim 20 \,\xi_v$, there is a low-density link of CdGM states between both vortices. This link can have significant implications regarding the energy dissipation, as shown already by the behavior in $\alpha$ in Supp. Fig. \ref{fig:extFig6}c. It could also give rise to a continuous exchange of quasiparticles between the vortices. These effects are not considered in the present analysis and are going to be presented in future works.

\begin{thebibliography}{76}%
\makeatletter
\providecommand \@ifxundefined [1]{%
 \@ifx{#1\undefined}
}%
\providecommand \@ifnum [1]{%
 \ifnum #1\expandafter \@firstoftwo
 \else \expandafter \@secondoftwo
 \fi
}%
\providecommand \@ifx [1]{%
 \ifx #1\expandafter \@firstoftwo
 \else \expandafter \@secondoftwo
 \fi
}%
\providecommand \natexlab [1]{#1}%
\providecommand \enquote  [1]{``#1''}%
\providecommand \bibnamefont  [1]{#1}%
\providecommand \bibfnamefont [1]{#1}%
\providecommand \citenamefont [1]{#1}%
\providecommand \href@noop [0]{\@secondoftwo}%
\providecommand \href [0]{\begingroup \@sanitize@url \@href}%
\providecommand \@href[1]{\@@startlink{#1}\@@href}%
\providecommand \@@href[1]{\endgroup#1\@@endlink}%
\providecommand \@sanitize@url [0]{\catcode `\\12\catcode `\$12\catcode
  `\&12\catcode `\#12\catcode `\^12\catcode `\_12\catcode `\%12\relax}%
\providecommand \@@startlink[1]{}%
\providecommand \@@endlink[0]{}%
\providecommand \url  [0]{\begingroup\@sanitize@url \@url }%
\providecommand \@url [1]{\endgroup\@href {#1}{\urlprefix }}%
\providecommand \urlprefix  [0]{URL }%
\providecommand \Eprint [0]{\href }%
\providecommand \doibase [0]{https://doi.org/}%
\providecommand \selectlanguage [0]{\@gobble}%
\providecommand \bibinfo  [0]{\@secondoftwo}%
\providecommand \bibfield  [0]{\@secondoftwo}%
\providecommand \translation [1]{[#1]}%
\providecommand \BibitemOpen [0]{}%
\providecommand \bibitemStop [0]{}%
\providecommand \bibitemNoStop [0]{.\EOS\space}%
\providecommand \EOS [0]{\spacefactor3000\relax}%
\providecommand \BibitemShut  [1]{\csname bibitem#1\endcsname}%
\let\auto@bib@innerbib\@empty
\bibitem [{\citenamefont {Smith}\ \emph {et~al.}(1993)\citenamefont {Smith},
  \citenamefont {Donnelly}, \citenamefont {Goldenfeld},\ and\ \citenamefont
  {Vinen}}]{smith1993decay}%
  \BibitemOpen
  \bibfield  {author} {\bibinfo {author} {\bibfnamefont {M.~R.}\ \bibnamefont
  {Smith}}, \bibinfo {author} {\bibfnamefont {R.~J.}\ \bibnamefont {Donnelly}},
  \bibinfo {author} {\bibfnamefont {N.}~\bibnamefont {Goldenfeld}},\ and\
  \bibinfo {author} {\bibfnamefont {W.}~\bibnamefont {Vinen}},\ }\bibfield
  {title} {\bibinfo {title} {Decay of vorticity in homogeneous turbulence},\
  }\href@noop {} {\bibfield  {journal} {\bibinfo  {journal} {Phys. Rev. Lett.}\
  }\textbf {\bibinfo {volume} {71}},\ \bibinfo {pages} {2583} (\bibinfo {year}
  {1993})}\BibitemShut {NoStop}%
\bibitem [{\citenamefont {Tsatsos}\ \emph {et~al.}(2016)\citenamefont
  {Tsatsos}, \citenamefont {Tavares}, \citenamefont {Cidrim}, \citenamefont
  {Fritsch}, \citenamefont {Caracanhas}, \citenamefont {dos Santos},
  \citenamefont {Barenghi},\ and\ \citenamefont
  {Bagnato}}]{tsatsos2016quantum}%
  \BibitemOpen
  \bibfield  {author} {\bibinfo {author} {\bibfnamefont {M.~C.}\ \bibnamefont
  {Tsatsos}}, \bibinfo {author} {\bibfnamefont {P.~E.}\ \bibnamefont
  {Tavares}}, \bibinfo {author} {\bibfnamefont {A.}~\bibnamefont {Cidrim}},
  \bibinfo {author} {\bibfnamefont {A.~R.}\ \bibnamefont {Fritsch}}, \bibinfo
  {author} {\bibfnamefont {M.~A.}\ \bibnamefont {Caracanhas}}, \bibinfo
  {author} {\bibfnamefont {F.~E.~A.}\ \bibnamefont {dos Santos}}, \bibinfo
  {author} {\bibfnamefont {C.~F.}\ \bibnamefont {Barenghi}},\ and\ \bibinfo
  {author} {\bibfnamefont {V.~S.}\ \bibnamefont {Bagnato}},\ }\bibfield
  {title} {\bibinfo {title} {Quantum turbulence in trapped atomic
  bose--einstein condensates},\ }\href@noop {} {\bibfield  {journal} {\bibinfo
  {journal} {Physics Reports}\ }\textbf {\bibinfo {volume} {622}},\ \bibinfo
  {pages} {1} (\bibinfo {year} {2016})}\BibitemShut {NoStop}%
\bibitem [{\citenamefont {Panico}\ \emph {et~al.}(2023)\citenamefont {Panico},
  \citenamefont {Comaron}, \citenamefont {Matuszewski}, \citenamefont
  {Lanotte}, \citenamefont {Trypogeorgos}, \citenamefont {Gigli}, \citenamefont
  {Giorgi}, \citenamefont {Ardizzone}, \citenamefont {Sanvitto},\ and\
  \citenamefont {Ballarini}}]{panico2023onset}%
  \BibitemOpen
  \bibfield  {author} {\bibinfo {author} {\bibfnamefont {R.}~\bibnamefont
  {Panico}}, \bibinfo {author} {\bibfnamefont {P.}~\bibnamefont {Comaron}},
  \bibinfo {author} {\bibfnamefont {M.}~\bibnamefont {Matuszewski}}, \bibinfo
  {author} {\bibfnamefont {A.~S.}\ \bibnamefont {Lanotte}}, \bibinfo {author}
  {\bibfnamefont {D.}~\bibnamefont {Trypogeorgos}}, \bibinfo {author}
  {\bibfnamefont {G.}~\bibnamefont {Gigli}}, \bibinfo {author} {\bibfnamefont
  {M.~D.}\ \bibnamefont {Giorgi}}, \bibinfo {author} {\bibfnamefont
  {V.}~\bibnamefont {Ardizzone}}, \bibinfo {author} {\bibfnamefont
  {D.}~\bibnamefont {Sanvitto}},\ and\ \bibinfo {author} {\bibfnamefont
  {D.}~\bibnamefont {Ballarini}},\ }\bibfield  {title} {\bibinfo {title} {Onset
  of vortex clustering and inverse energy cascade in dissipative quantum
  fluids},\ }\href {https://doi.org/10.1038/s41566-023-01174-4} {\bibfield
  {journal} {\bibinfo  {journal} {Nat. Phot.}\ }\textbf {\bibinfo {volume}
  {17}},\ \bibinfo {pages} {451} (\bibinfo {year} {2023})}\BibitemShut
  {NoStop}%
\bibitem [{\citenamefont {Huse}\ \emph {et~al.}(1992)\citenamefont {Huse},
  \citenamefont {Fisher},\ and\ \citenamefont
  {Fisher}}]{huse1992superconductors}%
  \BibitemOpen
  \bibfield  {author} {\bibinfo {author} {\bibfnamefont {D.~A.}\ \bibnamefont
  {Huse}}, \bibinfo {author} {\bibfnamefont {M.~P.}\ \bibnamefont {Fisher}},\
  and\ \bibinfo {author} {\bibfnamefont {D.~S.}\ \bibnamefont {Fisher}},\
  }\bibfield  {title} {\bibinfo {title} {Are superconductors really
  superconducting?},\ }\href@noop {} {\bibfield  {journal} {\bibinfo  {journal}
  {Nature}\ }\textbf {\bibinfo {volume} {358}},\ \bibinfo {pages} {553}
  (\bibinfo {year} {1992})}\BibitemShut {NoStop}%
\bibitem [{\citenamefont {Donnelly}\ and\ \citenamefont
  {Barenghi}(1998)}]{donnelly1998observed}%
  \BibitemOpen
  \bibfield  {author} {\bibinfo {author} {\bibfnamefont {R.~J.}\ \bibnamefont
  {Donnelly}}\ and\ \bibinfo {author} {\bibfnamefont {C.~F.}\ \bibnamefont
  {Barenghi}},\ }\bibfield  {title} {\bibinfo {title} {The observed properties
  of liquid helium at the saturated vapor pressure},\ }\href
  {https://srd.nist.gov/jpcrdreprint/1.556028.pdf} {\bibfield  {journal}
  {\bibinfo  {journal} {Journal of physical and chemical reference data}\
  }\textbf {\bibinfo {volume} {27}},\ \bibinfo {pages} {1217} (\bibinfo {year}
  {1998})}\BibitemShut {NoStop}%
\bibitem [{\citenamefont {Leggett}(2006)}]{Leggett2006-aw}%
  \BibitemOpen
  \bibfield  {author} {\bibinfo {author} {\bibfnamefont {A.~J.}\ \bibnamefont
  {Leggett}},\ }\href@noop {} {\emph {\bibinfo {title} {Quantum Liquids}}},\
  Oxford Graduate Texts\ (\bibinfo  {publisher} {Oxford University Press},\
  \bibinfo {address} {London, England},\ \bibinfo {year} {2006})\BibitemShut
  {NoStop}%
\bibitem [{\citenamefont {Halperin}\ \emph {et~al.}(2010)\citenamefont
  {Halperin}, \citenamefont {Refael},\ and\ \citenamefont
  {Demler}}]{Halperin2010}%
  \BibitemOpen
  \bibfield  {author} {\bibinfo {author} {\bibfnamefont {B.~I.}\ \bibnamefont
  {Halperin}}, \bibinfo {author} {\bibfnamefont {G.}~\bibnamefont {Refael}},\
  and\ \bibinfo {author} {\bibfnamefont {E.}~\bibnamefont {Demler}},\
  }\bibfield  {title} {\bibinfo {title} {Resistance in superconductors},\
  }\href {https://doi.org/10.1142/s021797921005644x} {\bibfield  {journal}
  {\bibinfo  {journal} {International Journal of Modern Physics B}\ }\textbf
  {\bibinfo {volume} {24}},\ \bibinfo {pages} {4039–4080} (\bibinfo {year}
  {2010})}\BibitemShut {NoStop}%
\bibitem [{\citenamefont {Anderson}\ and\ \citenamefont
  {Itoh}(1975)}]{anderson1975pulsar}%
  \BibitemOpen
  \bibfield  {author} {\bibinfo {author} {\bibfnamefont {P.}~\bibnamefont
  {Anderson}}\ and\ \bibinfo {author} {\bibfnamefont {N.}~\bibnamefont
  {Itoh}},\ }\bibfield  {title} {\bibinfo {title} {Pulsar glitches and
  restlessness as a hard superfluidity phenomenon},\ }\href@noop {} {\bibfield
  {journal} {\bibinfo  {journal} {Nature}\ }\textbf {\bibinfo {volume} {256}},\
  \bibinfo {pages} {25} (\bibinfo {year} {1975})}\BibitemShut {NoStop}%
\bibitem [{\citenamefont {Zhou}\ \emph {et~al.}(2022)\citenamefont {Zhou},
  \citenamefont {G{\"u}gercino{\u{g}}lu}, \citenamefont {Yuan}, \citenamefont
  {Ge},\ and\ \citenamefont {Yu}}]{zhou2022pulsar}%
  \BibitemOpen
  \bibfield  {author} {\bibinfo {author} {\bibfnamefont {S.}~\bibnamefont
  {Zhou}}, \bibinfo {author} {\bibfnamefont {E.}~\bibnamefont
  {G{\"u}gercino{\u{g}}lu}}, \bibinfo {author} {\bibfnamefont {J.}~\bibnamefont
  {Yuan}}, \bibinfo {author} {\bibfnamefont {M.}~\bibnamefont {Ge}},\ and\
  \bibinfo {author} {\bibfnamefont {C.}~\bibnamefont {Yu}},\ }\bibfield
  {title} {\bibinfo {title} {Pulsar glitches: A review},\ }\href@noop {}
  {\bibfield  {journal} {\bibinfo  {journal} {Universe}\ }\textbf {\bibinfo
  {volume} {8}},\ \bibinfo {pages} {641} (\bibinfo {year} {2022})}\BibitemShut
  {NoStop}%
\bibitem [{\citenamefont {Antonelli}\ and\ \citenamefont
  {Haskell}(2020)}]{antonelli2020superfluid}%
  \BibitemOpen
  \bibfield  {author} {\bibinfo {author} {\bibfnamefont {M.}~\bibnamefont
  {Antonelli}}\ and\ \bibinfo {author} {\bibfnamefont {B.}~\bibnamefont
  {Haskell}},\ }\bibfield  {title} {\bibinfo {title} {Superfluid
  vortex-mediated mutual friction in non-homogeneous neutron star interiors},\
  }\href@noop {} {\bibfield  {journal} {\bibinfo  {journal} {Monthly Notices of
  the Royal Astronomical Society}\ }\textbf {\bibinfo {volume} {499}},\
  \bibinfo {pages} {3690} (\bibinfo {year} {2020})}\BibitemShut {NoStop}%
\bibitem [{\citenamefont {E.}\ and\ \citenamefont {F.}(1956)}]{HallVinen}%
  \BibitemOpen
  \bibfield  {author} {\bibinfo {author} {\bibfnamefont {H.~H.}\ \bibnamefont
  {E.}}\ and\ \bibinfo {author} {\bibfnamefont {V.~W.}\ \bibnamefont {F.}},\
  }\bibfield  {title} {\bibinfo {title} {The rotation of liquid helium ii i.
  experiments on the propagation of second sound in uniformly rotating helium
  ii},\ }\href {https://doi.org/10.1098/rspa.1956.0214} {\bibfield  {journal}
  {\bibinfo  {journal} {Proceedings of the Royal Society of London. Series A.
  Mathematical and Physical Sciences}\ }\textbf {\bibinfo {volume} {238}},\
  \bibinfo {pages} {204–214} (\bibinfo {year} {1956})}\BibitemShut {NoStop}%
\bibitem [{\citenamefont {Kopnin}(2002)}]{kopnin2002vortex}%
  \BibitemOpen
  \bibfield  {author} {\bibinfo {author} {\bibfnamefont {N.~B.}\ \bibnamefont
  {Kopnin}},\ }\bibfield  {title} {\bibinfo {title} {Vortex dynamics and mutual
  friction in superconductors and fermi superfluids},\ }\href
  {https://iopscience.iop.org/article/10.1088/0034-4885/65/11/202} {\bibfield
  {journal} {\bibinfo  {journal} {Reports on Progress in Physics}\ }\textbf
  {\bibinfo {volume} {65}},\ \bibinfo {pages} {1633} (\bibinfo {year}
  {2002})}\BibitemShut {NoStop}%
\bibitem [{\citenamefont {Sonin}(2015)}]{Soninbook2015}%
  \BibitemOpen
  \bibfield  {author} {\bibinfo {author} {\bibfnamefont {E.~B.}\ \bibnamefont
  {Sonin}},\ }\href {https://doi.org/10.1017/cbo9781139047616} {\emph {\bibinfo
  {title} {Dynamics of Quantised Vortices in Superfluids}}}\ (\bibinfo
  {publisher} {Cambridge University Press},\ \bibinfo {year}
  {2015})\BibitemShut {NoStop}%
\bibitem [{\citenamefont {Caroli}\ \emph {et~al.}(1964)\citenamefont {Caroli},
  \citenamefont {De~Gennes},\ and\ \citenamefont {Matricon}}]{caroli1964bound}%
  \BibitemOpen
  \bibfield  {author} {\bibinfo {author} {\bibfnamefont {C.}~\bibnamefont
  {Caroli}}, \bibinfo {author} {\bibfnamefont {P.}~\bibnamefont {De~Gennes}},\
  and\ \bibinfo {author} {\bibfnamefont {J.}~\bibnamefont {Matricon}},\
  }\bibfield  {title} {\bibinfo {title} {Bound fermion states on a vortex line
  in a type ii superconductor},\ }\href@noop {} {\bibfield  {journal} {\bibinfo
   {journal} {Physics Letters}\ }\textbf {\bibinfo {volume} {9}},\ \bibinfo
  {pages} {307} (\bibinfo {year} {1964})}\BibitemShut {NoStop}%
\bibitem [{\citenamefont {Sensarma}\ \emph {et~al.}(2006)\citenamefont
  {Sensarma}, \citenamefont {Randeria},\ and\ \citenamefont
  {Ho}}]{sensarma2006vortices}%
  \BibitemOpen
  \bibfield  {author} {\bibinfo {author} {\bibfnamefont {R.}~\bibnamefont
  {Sensarma}}, \bibinfo {author} {\bibfnamefont {M.}~\bibnamefont {Randeria}},\
  and\ \bibinfo {author} {\bibfnamefont {T.-L.}\ \bibnamefont {Ho}},\
  }\bibfield  {title} {\bibinfo {title} {Vortices in superfluid fermi gases
  through the bec to bcs crossover},\ }\href@noop {} {\bibfield  {journal}
  {\bibinfo  {journal} {Phys. Rev. Lett.}\ }\textbf {\bibinfo {volume} {96}},\
  \bibinfo {pages} {090403} (\bibinfo {year} {2006})}\BibitemShut {NoStop}%
\bibitem [{\citenamefont {Bevan}\ \emph
  {et~al.}(1997{\natexlab{a}})\citenamefont {Bevan}, \citenamefont {Manninen},
  \citenamefont {Cook}, \citenamefont {Hook}, \citenamefont {Hall},
  \citenamefont {Vachaspati},\ and\ \citenamefont
  {Volovik}}]{bevan1997momentum}%
  \BibitemOpen
  \bibfield  {author} {\bibinfo {author} {\bibfnamefont {T.}~\bibnamefont
  {Bevan}}, \bibinfo {author} {\bibfnamefont {A.}~\bibnamefont {Manninen}},
  \bibinfo {author} {\bibfnamefont {J.}~\bibnamefont {Cook}}, \bibinfo {author}
  {\bibfnamefont {J.}~\bibnamefont {Hook}}, \bibinfo {author} {\bibfnamefont
  {H.}~\bibnamefont {Hall}}, \bibinfo {author} {\bibfnamefont {T.}~\bibnamefont
  {Vachaspati}},\ and\ \bibinfo {author} {\bibfnamefont {G.}~\bibnamefont
  {Volovik}},\ }\bibfield  {title} {\bibinfo {title} {Momentum creation by
  vortices in superfluid 3he as a model of primordial baryogenesis},\
  }\href@noop {} {\bibfield  {journal} {\bibinfo  {journal} {Nature}\ }\textbf
  {\bibinfo {volume} {386}},\ \bibinfo {pages} {689} (\bibinfo {year}
  {1997}{\natexlab{a}})}\BibitemShut {NoStop}%
\bibitem [{\citenamefont {Berthod}\ \emph {et~al.}(2017)\citenamefont
  {Berthod}, \citenamefont {Maggio-Aprile}, \citenamefont {Bru\'er},
  \citenamefont {Erb},\ and\ \citenamefont {Renner}}]{Berthod2017}%
  \BibitemOpen
  \bibfield  {author} {\bibinfo {author} {\bibfnamefont {C.}~\bibnamefont
  {Berthod}}, \bibinfo {author} {\bibfnamefont {I.}~\bibnamefont
  {Maggio-Aprile}}, \bibinfo {author} {\bibfnamefont {J.}~\bibnamefont
  {Bru\'er}}, \bibinfo {author} {\bibfnamefont {A.}~\bibnamefont {Erb}},\ and\
  \bibinfo {author} {\bibfnamefont {C.}~\bibnamefont {Renner}},\ }\bibfield
  {title} {\bibinfo {title} {Observation of caroli--de gennes--matricon vortex
  states in
  ${\mathrm{yba}}_{2}{\mathrm{cu}}_{3}{\mathrm{o}}_{7\ensuremath{-}\ensuremath{\delta}}$},\
  }\href {https://doi.org/10.1103/PhysRevLett.119.237001} {\bibfield  {journal}
  {\bibinfo  {journal} {Phys. Rev. Lett.}\ }\textbf {\bibinfo {volume} {119}},\
  \bibinfo {pages} {237001} (\bibinfo {year} {2017})}\BibitemShut {NoStop}%
\bibitem [{\citenamefont {Chen}\ \emph {et~al.}(2018)\citenamefont {Chen},
  \citenamefont {Chen}, \citenamefont {Yang}, \citenamefont {Du}, \citenamefont
  {Zhu}, \citenamefont {Wang},\ and\ \citenamefont {Wen}}]{Chen2018}%
  \BibitemOpen
  \bibfield  {author} {\bibinfo {author} {\bibfnamefont {M.}~\bibnamefont
  {Chen}}, \bibinfo {author} {\bibfnamefont {X.}~\bibnamefont {Chen}}, \bibinfo
  {author} {\bibfnamefont {H.}~\bibnamefont {Yang}}, \bibinfo {author}
  {\bibfnamefont {Z.}~\bibnamefont {Du}}, \bibinfo {author} {\bibfnamefont
  {X.}~\bibnamefont {Zhu}}, \bibinfo {author} {\bibfnamefont {E.}~\bibnamefont
  {Wang}},\ and\ \bibinfo {author} {\bibfnamefont {H.-H.}\ \bibnamefont
  {Wen}},\ }\bibfield  {title} {\bibinfo {title} {Discrete energy levels of
  caroli-de gennes-matricon states in quantum limit in
  fete${}_0.55$se${}_0.45$},\ }\href
  {https://doi.org/10.1038/s41467-018-03404-8} {\bibfield  {journal} {\bibinfo
  {journal} {Nat. Comm.}\ }\textbf {\bibinfo {volume} {9}},\ \bibinfo {pages}
  {970} (\bibinfo {year} {2018})}\BibitemShut {NoStop}%
\bibitem [{\citenamefont {Bevan}\ \emph
  {et~al.}(1997{\natexlab{b}})\citenamefont {Bevan}, \citenamefont {Manninen},
  \citenamefont {Cook}, \citenamefont {Alles}, \citenamefont {Hook},\ and\
  \citenamefont {Hall}}]{bevan1997vortex}%
  \BibitemOpen
  \bibfield  {author} {\bibinfo {author} {\bibfnamefont {T.}~\bibnamefont
  {Bevan}}, \bibinfo {author} {\bibfnamefont {A.}~\bibnamefont {Manninen}},
  \bibinfo {author} {\bibfnamefont {J.}~\bibnamefont {Cook}}, \bibinfo {author}
  {\bibfnamefont {H.}~\bibnamefont {Alles}}, \bibinfo {author} {\bibfnamefont
  {J.}~\bibnamefont {Hook}},\ and\ \bibinfo {author} {\bibfnamefont
  {H.}~\bibnamefont {Hall}},\ }\bibfield  {title} {\bibinfo {title} {Vortex
  mutual friction in superfluid 3 he},\ }\href@noop {} {\bibfield  {journal}
  {\bibinfo  {journal} {Journal of low temperature physics}\ }\textbf {\bibinfo
  {volume} {109}},\ \bibinfo {pages} {423} (\bibinfo {year}
  {1997}{\natexlab{b}})}\BibitemShut {NoStop}%
\bibitem [{\citenamefont {M{\"a}kinen}\ and\ \citenamefont
  {Eltsov}(2018)}]{makinen2018mutual}%
  \BibitemOpen
  \bibfield  {author} {\bibinfo {author} {\bibfnamefont {J.}~\bibnamefont
  {M{\"a}kinen}}\ and\ \bibinfo {author} {\bibfnamefont {V.}~\bibnamefont
  {Eltsov}},\ }\bibfield  {title} {\bibinfo {title} {Mutual friction in
  superfluid he-b 3 in the low-temperature regime},\ }\href@noop {} {\bibfield
  {journal} {\bibinfo  {journal} {Phys. Rev. B}\ }\textbf {\bibinfo {volume}
  {97}},\ \bibinfo {pages} {014527} (\bibinfo {year} {2018})}\BibitemShut
  {NoStop}%
\bibitem [{\citenamefont {Kopnin}\ and\ \citenamefont
  {Salomaa}(1991)}]{kopnin1991mutual}%
  \BibitemOpen
  \bibfield  {author} {\bibinfo {author} {\bibfnamefont {N.}~\bibnamefont
  {Kopnin}}\ and\ \bibinfo {author} {\bibfnamefont {M.}~\bibnamefont
  {Salomaa}},\ }\bibfield  {title} {\bibinfo {title} {Mutual friction in
  superfluid he 3: Effects of bound states in the vortex core},\ }\href@noop {}
  {\bibfield  {journal} {\bibinfo  {journal} {Phys. Rev. B}\ }\textbf {\bibinfo
  {volume} {44}},\ \bibinfo {pages} {9667} (\bibinfo {year}
  {1991})}\BibitemShut {NoStop}%
\bibitem [{\citenamefont {Ku}\ \emph {et~al.}(2014)\citenamefont {Ku},
  \citenamefont {Ji}, \citenamefont {Mukherjee}, \citenamefont
  {Guardado-Sanchez}, \citenamefont {Cheuk}, \citenamefont {Yefsah},\ and\
  \citenamefont {Zwierlein}}]{ku2014motion}%
  \BibitemOpen
  \bibfield  {author} {\bibinfo {author} {\bibfnamefont {M.~J.}\ \bibnamefont
  {Ku}}, \bibinfo {author} {\bibfnamefont {W.}~\bibnamefont {Ji}}, \bibinfo
  {author} {\bibfnamefont {B.}~\bibnamefont {Mukherjee}}, \bibinfo {author}
  {\bibfnamefont {E.}~\bibnamefont {Guardado-Sanchez}}, \bibinfo {author}
  {\bibfnamefont {L.~W.}\ \bibnamefont {Cheuk}}, \bibinfo {author}
  {\bibfnamefont {T.}~\bibnamefont {Yefsah}},\ and\ \bibinfo {author}
  {\bibfnamefont {M.~W.}\ \bibnamefont {Zwierlein}},\ }\bibfield  {title}
  {\bibinfo {title} {Motion of a solitonic vortex in the bec-bcs crossover},\
  }\href@noop {} {\bibfield  {journal} {\bibinfo  {journal} {Phys. Rev. Lett.}\
  }\textbf {\bibinfo {volume} {113}},\ \bibinfo {pages} {065301} (\bibinfo
  {year} {2014})}\BibitemShut {NoStop}%
\bibitem [{\citenamefont {Park}\ \emph {et~al.}(2018)\citenamefont {Park},
  \citenamefont {Ko},\ and\ \citenamefont {Shin}}]{park2018critical}%
  \BibitemOpen
  \bibfield  {author} {\bibinfo {author} {\bibfnamefont {J.~W.}\ \bibnamefont
  {Park}}, \bibinfo {author} {\bibfnamefont {B.}~\bibnamefont {Ko}},\ and\
  \bibinfo {author} {\bibfnamefont {Y.-i.}\ \bibnamefont {Shin}},\ }\bibfield
  {title} {\bibinfo {title} {Critical vortex shedding in a strongly interacting
  fermionic superfluid},\ }\href@noop {} {\bibfield  {journal} {\bibinfo
  {journal} {Phys. Rev. Lett.}\ }\textbf {\bibinfo {volume} {121}},\ \bibinfo
  {pages} {225301} (\bibinfo {year} {2018})}\BibitemShut {NoStop}%
\bibitem [{\citenamefont {Kwon}\ \emph {et~al.}(2021)\citenamefont {Kwon} \emph
  {et~al.}}]{kwon2021sound}%
  \BibitemOpen
  \bibfield  {author} {\bibinfo {author} {\bibfnamefont {W.~J.}\ \bibnamefont
  {Kwon}} \emph {et~al.},\ }\bibfield  {title} {\bibinfo {title} {Sound
  emission and annihilations in a programmable quantum vortex collider},\
  }\href {https://www.nature.com/articles/s41586-021-04047-4} {\bibfield
  {journal} {\bibinfo  {journal} {Nature}\ }\textbf {\bibinfo {volume} {600}},\
  \bibinfo {pages} {64} (\bibinfo {year} {2021})}\BibitemShut {NoStop}%
\bibitem [{\citenamefont {Bennemann}\ and\ \citenamefont
  {Ketterson}(2014)}]{novelSF2}%
  \BibitemOpen
  \bibfield  {author} {\bibinfo {author} {\bibfnamefont {K.-H.}\ \bibnamefont
  {Bennemann}}\ and\ \bibinfo {author} {\bibfnamefont {J.~B.}\ \bibnamefont
  {Ketterson}},\ }\href
  {https://doi.org/10.1093/acprof:oso/9780198719267.001.0001} {\emph {\bibinfo
  {title} {{Novel Superfluids: Volume 2}}}}\ (\bibinfo  {publisher} {Oxford
  University Press},\ \bibinfo {year} {2014})\BibitemShut {NoStop}%
\bibitem [{\citenamefont {Barresi}\ \emph {et~al.}(2023)\citenamefont
  {Barresi}, \citenamefont {Boulet}, \citenamefont {Magierski},\ and\
  \citenamefont {Wlaz{\l}owski}}]{barresi2023dissipative}%
  \BibitemOpen
  \bibfield  {author} {\bibinfo {author} {\bibfnamefont {A.}~\bibnamefont
  {Barresi}}, \bibinfo {author} {\bibfnamefont {A.}~\bibnamefont {Boulet}},
  \bibinfo {author} {\bibfnamefont {P.}~\bibnamefont {Magierski}},\ and\
  \bibinfo {author} {\bibfnamefont {G.}~\bibnamefont {Wlaz{\l}owski}},\
  }\bibfield  {title} {\bibinfo {title} {Dissipative dynamics of quantum
  vortices in fermionic superfluid},\ }\href@noop {} {\bibfield  {journal}
  {\bibinfo  {journal} {Phys. Rev. Lett.}\ }\textbf {\bibinfo {volume} {130}},\
  \bibinfo {pages} {043001} (\bibinfo {year} {2023})}\BibitemShut {NoStop}%
\bibitem [{\citenamefont {Schwarz}(1985)}]{schwarz1985three}%
  \BibitemOpen
  \bibfield  {author} {\bibinfo {author} {\bibfnamefont {K.}~\bibnamefont
  {Schwarz}},\ }\bibfield  {title} {\bibinfo {title} {Three-dimensional vortex
  dynamics in superfluid he 4: Line-line and line-boundary interactions},\
  }\href@noop {} {\bibfield  {journal} {\bibinfo  {journal} {Phys. Rev. B}\
  }\textbf {\bibinfo {volume} {31}},\ \bibinfo {pages} {5782} (\bibinfo {year}
  {1985})}\BibitemShut {NoStop}%
\bibitem [{\citenamefont {Sergeev}(2023)}]{Sergeev2023}%
  \BibitemOpen
  \bibfield  {author} {\bibinfo {author} {\bibfnamefont {Y.~A.}\ \bibnamefont
  {Sergeev}},\ }\bibfield  {title} {\bibinfo {title} {Mutual friction in
  bosonic superfluids: A review},\ }\href
  {https://doi.org/10.1007/s10909-023-02972-4} {\bibfield  {journal} {\bibinfo
  {journal} {Journal of Low Temperature Physics}\ }\textbf {\bibinfo {volume}
  {212}},\ \bibinfo {pages} {251–305} (\bibinfo {year} {2023})}\BibitemShut
  {NoStop}%
\bibitem [{\citenamefont {Minowa}\ \emph {et~al.}(2025)\citenamefont {Minowa},
  \citenamefont {Yasui}, \citenamefont {Nakagawa}, \citenamefont {Inui},
  \citenamefont {Tsubota},\ and\ \citenamefont {Ashida}}]{minowa2025direct}%
  \BibitemOpen
  \bibfield  {author} {\bibinfo {author} {\bibfnamefont {Y.}~\bibnamefont
  {Minowa}}, \bibinfo {author} {\bibfnamefont {Y.}~\bibnamefont {Yasui}},
  \bibinfo {author} {\bibfnamefont {T.}~\bibnamefont {Nakagawa}}, \bibinfo
  {author} {\bibfnamefont {S.}~\bibnamefont {Inui}}, \bibinfo {author}
  {\bibfnamefont {M.}~\bibnamefont {Tsubota}},\ and\ \bibinfo {author}
  {\bibfnamefont {M.}~\bibnamefont {Ashida}},\ }\bibfield  {title} {\bibinfo
  {title} {Direct excitation of kelvin waves on quantized vortices},\ }\href
  {https://doi.org/10.1038/s41567-024-02720-9} {\bibfield  {journal} {\bibinfo
  {journal} {Nature Physics}\ }\textbf {\bibinfo {volume} {21}},\ \bibinfo
  {pages} {233–238} (\bibinfo {year} {2025})}\BibitemShut {NoStop}%
\bibitem [{\citenamefont {Tang}\ \emph {et~al.}(2023)\citenamefont {Tang},
  \citenamefont {Guo}, \citenamefont {Kobayashi}, \citenamefont {Yui},
  \citenamefont {Tsubota},\ and\ \citenamefont {Kanai}}]{tang2023imaging}%
  \BibitemOpen
  \bibfield  {author} {\bibinfo {author} {\bibfnamefont {Y.}~\bibnamefont
  {Tang}}, \bibinfo {author} {\bibfnamefont {W.}~\bibnamefont {Guo}}, \bibinfo
  {author} {\bibfnamefont {H.}~\bibnamefont {Kobayashi}}, \bibinfo {author}
  {\bibfnamefont {S.}~\bibnamefont {Yui}}, \bibinfo {author} {\bibfnamefont
  {M.}~\bibnamefont {Tsubota}},\ and\ \bibinfo {author} {\bibfnamefont
  {T.}~\bibnamefont {Kanai}},\ }\bibfield  {title} {\bibinfo {title} {Imaging
  quantized vortex rings in superfluid helium to evaluate quantum
  dissipation},\ }\href@noop {} {\bibfield  {journal} {\bibinfo  {journal}
  {Nature communications}\ }\textbf {\bibinfo {volume} {14}},\ \bibinfo {pages}
  {2941} (\bibinfo {year} {2023})}\BibitemShut {NoStop}%
\bibitem [{\citenamefont {Moon}\ \emph {et~al.}(2015)\citenamefont {Moon},
  \citenamefont {Kwon}, \citenamefont {Lee},\ and\ \citenamefont
  {Shin}}]{moon2015thermal}%
  \BibitemOpen
  \bibfield  {author} {\bibinfo {author} {\bibfnamefont {G.}~\bibnamefont
  {Moon}}, \bibinfo {author} {\bibfnamefont {W.~J.}\ \bibnamefont {Kwon}},
  \bibinfo {author} {\bibfnamefont {H.}~\bibnamefont {Lee}},\ and\ \bibinfo
  {author} {\bibfnamefont {Y.-i.}\ \bibnamefont {Shin}},\ }\bibfield  {title}
  {\bibinfo {title} {Thermal friction on quantum vortices in a bose-einstein
  condensate},\ }\href@noop {} {\bibfield  {journal} {\bibinfo  {journal}
  {Phys. Rev. A}\ }\textbf {\bibinfo {volume} {92}},\ \bibinfo {pages} {051601}
  (\bibinfo {year} {2015})}\BibitemShut {NoStop}%
\bibitem [{\citenamefont {Kim}\ \emph {et~al.}(2016)\citenamefont {Kim},
  \citenamefont {Kwon},\ and\ \citenamefont {Shin}}]{kim2016role}%
  \BibitemOpen
  \bibfield  {author} {\bibinfo {author} {\bibfnamefont {J.~H.}\ \bibnamefont
  {Kim}}, \bibinfo {author} {\bibfnamefont {W.~J.}\ \bibnamefont {Kwon}},\ and\
  \bibinfo {author} {\bibfnamefont {Y.-i.}\ \bibnamefont {Shin}},\ }\bibfield
  {title} {\bibinfo {title} {Role of thermal friction in relaxation of
  turbulent bose-einstein condensates},\ }\href@noop {} {\bibfield  {journal}
  {\bibinfo  {journal} {Phys. Rev. A}\ }\textbf {\bibinfo {volume} {94}},\
  \bibinfo {pages} {033612} (\bibinfo {year} {2016})}\BibitemShut {NoStop}%
\bibitem [{\citenamefont {Neely}\ \emph {et~al.}(2024)\citenamefont {Neely},
  \citenamefont {Gauthier}, \citenamefont {Glasspool}, \citenamefont {Davis},\
  and\ \citenamefont {Reeves}}]{neely2024melting}%
  \BibitemOpen
  \bibfield  {author} {\bibinfo {author} {\bibfnamefont {T.~W.}\ \bibnamefont
  {Neely}}, \bibinfo {author} {\bibfnamefont {G.}~\bibnamefont {Gauthier}},
  \bibinfo {author} {\bibfnamefont {C.}~\bibnamefont {Glasspool}}, \bibinfo
  {author} {\bibfnamefont {M.~J.}\ \bibnamefont {Davis}},\ and\ \bibinfo
  {author} {\bibfnamefont {M.~T.}\ \bibnamefont {Reeves}},\ }\bibfield  {title}
  {\bibinfo {title} {Melting of a vortex matter wigner crystal},\ }\href@noop
  {} {\bibfield  {journal} {\bibinfo  {journal} {arXiv preprint
  arXiv:2402.09920}\ } (\bibinfo {year} {2024})}\BibitemShut {NoStop}%
\bibitem [{\citenamefont {Bulgac}(2007)}]{Bulgac2007}%
  \BibitemOpen
  \bibfield  {author} {\bibinfo {author} {\bibfnamefont {A.}~\bibnamefont
  {Bulgac}},\ }\bibfield  {title} {\bibinfo {title} {Local-density-functional
  theory for superfluid fermionic systems: The unitary gas},\ }\href
  {https://doi.org/10.1103/PhysRevA.76.040502} {\bibfield  {journal} {\bibinfo
  {journal} {Phys. Rev. A}\ }\textbf {\bibinfo {volume} {76}},\ \bibinfo
  {pages} {040502} (\bibinfo {year} {2007})}\BibitemShut {NoStop}%
\bibitem [{\citenamefont {Heyl}\ \emph {et~al.}(2022)\citenamefont {Heyl},
  \citenamefont {Adachi}, \citenamefont {Itahashi}, \citenamefont {Nakagawa},
  \citenamefont {Kasahara}, \citenamefont {List-Kratochvil}, \citenamefont
  {Kato},\ and\ \citenamefont {Iwasa}}]{heyl2022vortex}%
  \BibitemOpen
  \bibfield  {author} {\bibinfo {author} {\bibfnamefont {M.}~\bibnamefont
  {Heyl}}, \bibinfo {author} {\bibfnamefont {K.}~\bibnamefont {Adachi}},
  \bibinfo {author} {\bibfnamefont {Y.~M.}\ \bibnamefont {Itahashi}}, \bibinfo
  {author} {\bibfnamefont {Y.}~\bibnamefont {Nakagawa}}, \bibinfo {author}
  {\bibfnamefont {Y.}~\bibnamefont {Kasahara}}, \bibinfo {author}
  {\bibfnamefont {E.~J.}\ \bibnamefont {List-Kratochvil}}, \bibinfo {author}
  {\bibfnamefont {Y.}~\bibnamefont {Kato}},\ and\ \bibinfo {author}
  {\bibfnamefont {Y.}~\bibnamefont {Iwasa}},\ }\bibfield  {title} {\bibinfo
  {title} {Vortex dynamics in the two-dimensional bcs-bec crossover},\
  }\href@noop {} {\bibfield  {journal} {\bibinfo  {journal} {Nature
  communications}\ }\textbf {\bibinfo {volume} {13}},\ \bibinfo {pages} {6986}
  (\bibinfo {year} {2022})}\BibitemShut {NoStop}%
\bibitem [{\citenamefont {Ogawa}\ \emph {et~al.}(2021)\citenamefont {Ogawa},
  \citenamefont {Nabeshima}, \citenamefont {Nishizaki},\ and\ \citenamefont
  {Maeda}}]{ogawa2021large}%
  \BibitemOpen
  \bibfield  {author} {\bibinfo {author} {\bibfnamefont {R.}~\bibnamefont
  {Ogawa}}, \bibinfo {author} {\bibfnamefont {F.}~\bibnamefont {Nabeshima}},
  \bibinfo {author} {\bibfnamefont {T.}~\bibnamefont {Nishizaki}},\ and\
  \bibinfo {author} {\bibfnamefont {A.}~\bibnamefont {Maeda}},\ }\bibfield
  {title} {\bibinfo {title} {Large hall angle of vortex motion in high-t c
  cuprate superconductors revealed by microwave flux-flow hall effect},\
  }\href@noop {} {\bibfield  {journal} {\bibinfo  {journal} {Phys. Rev. B}\
  }\textbf {\bibinfo {volume} {104}},\ \bibinfo {pages} {L020503} (\bibinfo
  {year} {2021})}\BibitemShut {NoStop}%
\bibitem [{\citenamefont {Finne}\ \emph {et~al.}(2003)\citenamefont {Finne},
  \citenamefont {Araki}, \citenamefont {Blaauwgeers}, \citenamefont {Eltsov},
  \citenamefont {Kopnin}, \citenamefont {Krusius}, \citenamefont {Skrbek},
  \citenamefont {Tsubota},\ and\ \citenamefont {Volovik}}]{finne2003intrinsic}%
  \BibitemOpen
  \bibfield  {author} {\bibinfo {author} {\bibfnamefont {A.}~\bibnamefont
  {Finne}}, \bibinfo {author} {\bibfnamefont {T.}~\bibnamefont {Araki}},
  \bibinfo {author} {\bibfnamefont {R.}~\bibnamefont {Blaauwgeers}}, \bibinfo
  {author} {\bibfnamefont {V.}~\bibnamefont {Eltsov}}, \bibinfo {author}
  {\bibfnamefont {N.}~\bibnamefont {Kopnin}}, \bibinfo {author} {\bibfnamefont
  {M.}~\bibnamefont {Krusius}}, \bibinfo {author} {\bibfnamefont
  {L.}~\bibnamefont {Skrbek}}, \bibinfo {author} {\bibfnamefont
  {M.}~\bibnamefont {Tsubota}},\ and\ \bibinfo {author} {\bibfnamefont
  {G.}~\bibnamefont {Volovik}},\ }\bibfield  {title} {\bibinfo {title} {An
  intrinsic velocity-independent criterion for superfluid turbulence},\
  }\href@noop {} {\bibfield  {journal} {\bibinfo  {journal} {Nature}\ }\textbf
  {\bibinfo {volume} {424}},\ \bibinfo {pages} {1022} (\bibinfo {year}
  {2003})}\BibitemShut {NoStop}%
\bibitem [{\citenamefont {Hueck}\ \emph {et~al.}(2018)\citenamefont {Hueck}
  \emph {et~al.}}]{Temp_Measure_Moritz}%
  \BibitemOpen
  \bibfield  {author} {\bibinfo {author} {\bibfnamefont {K.}~\bibnamefont
  {Hueck}} \emph {et~al.},\ }\bibfield  {title} {\bibinfo {title}
  {Two-dimensional homogeneous fermi gases},\ }\href
  {https://doi.org/10.1103/PhysRevLett.120.060402} {\bibfield  {journal}
  {\bibinfo  {journal} {Phys. Rev. Lett.}\ }\textbf {\bibinfo {volume} {120}},\
  \bibinfo {pages} {060402} (\bibinfo {year} {2018})}\BibitemShut {NoStop}%
\bibitem [{\citenamefont {Pini}\ \emph {et~al.}(2019)\citenamefont {Pini},
  \citenamefont {Pieri},\ and\ \citenamefont {Strinati}}]{Pini2019}%
  \BibitemOpen
  \bibfield  {author} {\bibinfo {author} {\bibfnamefont {M.}~\bibnamefont
  {Pini}}, \bibinfo {author} {\bibfnamefont {P.}~\bibnamefont {Pieri}},\ and\
  \bibinfo {author} {\bibfnamefont {G.~C.}\ \bibnamefont {Strinati}},\
  }\bibfield  {title} {\bibinfo {title} {Fermi gas throughout the bcs-bec
  crossover: Comparative study of $t$-matrix approaches with various degrees of
  self-consistency},\ }\href {https://doi.org/10.1103/PhysRevB.99.094502}
  {\bibfield  {journal} {\bibinfo  {journal} {Phys. Rev. B}\ }\textbf {\bibinfo
  {volume} {99}},\ \bibinfo {pages} {094502} (\bibinfo {year}
  {2019})}\BibitemShut {NoStop}%
\bibitem [{\citenamefont {Samson}\ \emph {et~al.}(2016)\citenamefont {Samson}
  \emph {et~al.}}]{Samson_DeterministicCreation}%
  \BibitemOpen
  \bibfield  {author} {\bibinfo {author} {\bibfnamefont {E.~C.}\ \bibnamefont
  {Samson}} \emph {et~al.},\ }\bibfield  {title} {\bibinfo {title}
  {Deterministic creation, pinning, and manipulation of quantized vortices in a
  bose-einstein condensate},\ }\href
  {https://doi.org/10.1103/PhysRevA.93.023603} {\bibfield  {journal} {\bibinfo
  {journal} {Phys. Rev. A}\ }\textbf {\bibinfo {volume} {93}},\ \bibinfo
  {pages} {023603} (\bibinfo {year} {2016})}\BibitemShut {NoStop}%
\bibitem [{\citenamefont {Vinen}(2000)}]{vinen2000classical}%
  \BibitemOpen
  \bibfield  {author} {\bibinfo {author} {\bibfnamefont {W.}~\bibnamefont
  {Vinen}},\ }\bibfield  {title} {\bibinfo {title} {Classical character of
  turbulence in a quantum liquid},\ }\href@noop {} {\bibfield  {journal}
  {\bibinfo  {journal} {Phys. Rev. B}\ }\textbf {\bibinfo {volume} {61}},\
  \bibinfo {pages} {1410} (\bibinfo {year} {2000})}\BibitemShut {NoStop}%
\bibitem [{\citenamefont {Simjanovski}\ \emph {et~al.}(2024)\citenamefont
  {Simjanovski}, \citenamefont {Gauthier}, \citenamefont {Rubinsztein-Dunlop},
  \citenamefont {Reeves},\ and\ \citenamefont {Neely}}]{simjanovski2024shear}%
  \BibitemOpen
  \bibfield  {author} {\bibinfo {author} {\bibfnamefont {S.}~\bibnamefont
  {Simjanovski}}, \bibinfo {author} {\bibfnamefont {G.}~\bibnamefont
  {Gauthier}}, \bibinfo {author} {\bibfnamefont {H.}~\bibnamefont
  {Rubinsztein-Dunlop}}, \bibinfo {author} {\bibfnamefont {M.~T.}\ \bibnamefont
  {Reeves}},\ and\ \bibinfo {author} {\bibfnamefont {T.~W.}\ \bibnamefont
  {Neely}},\ }\bibfield  {title} {\bibinfo {title} {Shear-induced decaying
  turbulence in bose-einstein condensates},\ }\href@noop {} {\bibfield
  {journal} {\bibinfo  {journal} {arXiv preprint arXiv:2408.02200}\ } (\bibinfo
  {year} {2024})}\BibitemShut {NoStop}%
\bibitem [{\citenamefont {Jackson}\ \emph {et~al.}(2009)\citenamefont
  {Jackson}, \citenamefont {Proukakis}, \citenamefont {Barenghi},\ and\
  \citenamefont {Zaremba}}]{Jackson2009}%
  \BibitemOpen
  \bibfield  {author} {\bibinfo {author} {\bibfnamefont {B.}~\bibnamefont
  {Jackson}}, \bibinfo {author} {\bibfnamefont {N.~P.}\ \bibnamefont
  {Proukakis}}, \bibinfo {author} {\bibfnamefont {C.~F.}\ \bibnamefont
  {Barenghi}},\ and\ \bibinfo {author} {\bibfnamefont {E.}~\bibnamefont
  {Zaremba}},\ }\bibfield  {title} {\bibinfo {title} {Finite-temperature vortex
  dynamics in bose-einstein condensates},\ }\bibfield  {journal} {\bibinfo
  {journal} {Phys. Rev. A}\ }\textbf {\bibinfo {volume} {79}},\ \href
  {https://doi.org/10.1103/physreva.79.053615} {10.1103/physreva.79.053615}
  (\bibinfo {year} {2009})\BibitemShut {NoStop}%
\bibitem [{\citenamefont {Mozyrsky}\ and\ \citenamefont
  {Chubukov}(2019)}]{Mozyrsky2019}%
  \BibitemOpen
  \bibfield  {author} {\bibinfo {author} {\bibfnamefont {D.}~\bibnamefont
  {Mozyrsky}}\ and\ \bibinfo {author} {\bibfnamefont {A.~V.}\ \bibnamefont
  {Chubukov}},\ }\bibfield  {title} {\bibinfo {title} {Dynamic properties of
  superconductors: Anderson-bogoliubov mode and berry phase in the bcs and bec
  regimes},\ }\bibfield  {journal} {\bibinfo  {journal} {Phys. Rev. B}\
  }\textbf {\bibinfo {volume} {99}},\ \href
  {https://doi.org/10.1103/physrevb.99.174510} {10.1103/physrevb.99.174510}
  (\bibinfo {year} {2019})\BibitemShut {NoStop}%
\bibitem [{\citenamefont {Pitaevskii}(1959)}]{pitaevskii1959calculation}%
  \BibitemOpen
  \bibfield  {author} {\bibinfo {author} {\bibfnamefont {L.}~\bibnamefont
  {Pitaevskii}},\ }\bibfield  {title} {\bibinfo {title} {Calculation of the
  phonon part of the mutual friction force in superfluid helium},\ }\href@noop
  {} {\bibfield  {journal} {\bibinfo  {journal} {Sov. Phys. JETP}\ }\textbf
  {\bibinfo {volume} {8}},\ \bibinfo {pages} {888} (\bibinfo {year}
  {1959})}\BibitemShut {NoStop}%
\bibitem [{\citenamefont {Barenghi}\ \emph {et~al.}(1983)\citenamefont
  {Barenghi}, \citenamefont {Donnelly},\ and\ \citenamefont
  {Vinen}}]{barenghi1983friction}%
  \BibitemOpen
  \bibfield  {author} {\bibinfo {author} {\bibfnamefont {C.}~\bibnamefont
  {Barenghi}}, \bibinfo {author} {\bibfnamefont {R.}~\bibnamefont {Donnelly}},\
  and\ \bibinfo {author} {\bibfnamefont {W.}~\bibnamefont {Vinen}},\ }\bibfield
   {title} {\bibinfo {title} {Friction on quantized vortices in helium ii. a
  review},\ }\href@noop {} {\bibfield  {journal} {\bibinfo  {journal} {Journal
  of Low Temperature Physics}\ }\textbf {\bibinfo {volume} {52}},\ \bibinfo
  {pages} {189} (\bibinfo {year} {1983})}\BibitemShut {NoStop}%
\bibitem [{\citenamefont {Silaev}(2012)}]{silaev2012universal}%
  \BibitemOpen
  \bibfield  {author} {\bibinfo {author} {\bibfnamefont {M.~A.}\ \bibnamefont
  {Silaev}},\ }\bibfield  {title} {\bibinfo {title} {Universal mechanism of
  dissipation in fermi superfluids at ultralow temperatures},\ }\href@noop {}
  {\bibfield  {journal} {\bibinfo  {journal} {Phys. Rev. Lett.}\ }\textbf
  {\bibinfo {volume} {108}},\ \bibinfo {pages} {045303} (\bibinfo {year}
  {2012})}\BibitemShut {NoStop}%
\bibitem [{\citenamefont {Eltsov}\ \emph {et~al.}(2007)\citenamefont {Eltsov},
  \citenamefont {Golov}, \citenamefont {de~Graaf}, \citenamefont {H\"anninen},
  \citenamefont {Krusius}, \citenamefont {L'vov},\ and\ \citenamefont
  {Solntsev}}]{eltsov2007}%
  \BibitemOpen
  \bibfield  {author} {\bibinfo {author} {\bibfnamefont {V.~B.}\ \bibnamefont
  {Eltsov}}, \bibinfo {author} {\bibfnamefont {A.~I.}\ \bibnamefont {Golov}},
  \bibinfo {author} {\bibfnamefont {R.}~\bibnamefont {de~Graaf}}, \bibinfo
  {author} {\bibfnamefont {R.}~\bibnamefont {H\"anninen}}, \bibinfo {author}
  {\bibfnamefont {M.}~\bibnamefont {Krusius}}, \bibinfo {author} {\bibfnamefont
  {V.~S.}\ \bibnamefont {L'vov}},\ and\ \bibinfo {author} {\bibfnamefont
  {R.~E.}\ \bibnamefont {Solntsev}},\ }\bibfield  {title} {\bibinfo {title}
  {Quantum turbulence in a propagating superfluid vortex front},\ }\href
  {https://doi.org/10.1103/PhysRevLett.99.265301} {\bibfield  {journal}
  {\bibinfo  {journal} {Phys. Rev. Lett.}\ }\textbf {\bibinfo {volume} {99}},\
  \bibinfo {pages} {265301} (\bibinfo {year} {2007})}\BibitemShut {NoStop}%
\bibitem [{\citenamefont {Yan}\ \emph {et~al.}(2024)\citenamefont {Yan},
  \citenamefont {Patel}, \citenamefont {Mukherjee}, \citenamefont {Vale},
  \citenamefont {Fletcher},\ and\ \citenamefont
  {Zwierlein}}]{yan2024thermography}%
  \BibitemOpen
  \bibfield  {author} {\bibinfo {author} {\bibfnamefont {Z.}~\bibnamefont
  {Yan}}, \bibinfo {author} {\bibfnamefont {P.~B.}\ \bibnamefont {Patel}},
  \bibinfo {author} {\bibfnamefont {B.}~\bibnamefont {Mukherjee}}, \bibinfo
  {author} {\bibfnamefont {C.~J.}\ \bibnamefont {Vale}}, \bibinfo {author}
  {\bibfnamefont {R.~J.}\ \bibnamefont {Fletcher}},\ and\ \bibinfo {author}
  {\bibfnamefont {M.~W.}\ \bibnamefont {Zwierlein}},\ }\bibfield  {title}
  {\bibinfo {title} {Thermography of the superfluid transition in a strongly
  interacting fermi gas},\ }\href@noop {} {\bibfield  {journal} {\bibinfo
  {journal} {Science}\ }\textbf {\bibinfo {volume} {383}},\ \bibinfo {pages}
  {629} (\bibinfo {year} {2024})}\BibitemShut {NoStop}%
\bibitem [{\citenamefont {Li}\ \emph {et~al.}(2022)\citenamefont {Li},
  \citenamefont {Luo}, \citenamefont {Wang}, \citenamefont {Xie}, \citenamefont
  {Liu}, \citenamefont {Hu}, \citenamefont {Chen}, \citenamefont {Yao},\ and\
  \citenamefont {Pan}}]{li2022second}%
  \BibitemOpen
  \bibfield  {author} {\bibinfo {author} {\bibfnamefont {X.}~\bibnamefont
  {Li}}, \bibinfo {author} {\bibfnamefont {X.}~\bibnamefont {Luo}}, \bibinfo
  {author} {\bibfnamefont {S.}~\bibnamefont {Wang}}, \bibinfo {author}
  {\bibfnamefont {K.}~\bibnamefont {Xie}}, \bibinfo {author} {\bibfnamefont
  {X.-P.}\ \bibnamefont {Liu}}, \bibinfo {author} {\bibfnamefont
  {H.}~\bibnamefont {Hu}}, \bibinfo {author} {\bibfnamefont {Y.-A.}\
  \bibnamefont {Chen}}, \bibinfo {author} {\bibfnamefont {X.-C.}\ \bibnamefont
  {Yao}},\ and\ \bibinfo {author} {\bibfnamefont {J.-W.}\ \bibnamefont {Pan}},\
  }\bibfield  {title} {\bibinfo {title} {Second sound attenuation near quantum
  criticality},\ }\href@noop {} {\bibfield  {journal} {\bibinfo  {journal}
  {Science}\ }\textbf {\bibinfo {volume} {375}},\ \bibinfo {pages} {528}
  (\bibinfo {year} {2022})}\BibitemShut {NoStop}%
\bibitem [{\citenamefont {Barenghi}\ \emph {et~al.}(1985)\citenamefont
  {Barenghi}, \citenamefont {Donnelly},\ and\ \citenamefont
  {Vinen}}]{barenghi1985thermal}%
  \BibitemOpen
  \bibfield  {author} {\bibinfo {author} {\bibfnamefont {C.}~\bibnamefont
  {Barenghi}}, \bibinfo {author} {\bibfnamefont {R.~J.}\ \bibnamefont
  {Donnelly}},\ and\ \bibinfo {author} {\bibfnamefont {W.}~\bibnamefont
  {Vinen}},\ }\bibfield  {title} {\bibinfo {title} {Thermal excitation of waves
  on quantized vortices},\ }\href@noop {} {\bibfield  {journal} {\bibinfo
  {journal} {The Physics of fluids}\ }\textbf {\bibinfo {volume} {28}},\
  \bibinfo {pages} {498} (\bibinfo {year} {1985})}\BibitemShut {NoStop}%
\bibitem [{\citenamefont {Volovik}(2003)}]{volovik2003classical}%
  \BibitemOpen
  \bibfield  {author} {\bibinfo {author} {\bibfnamefont {G.~E.}\ \bibnamefont
  {Volovik}},\ }\bibfield  {title} {\bibinfo {title} {Classical and quantum
  regimes of superfluid turbulence},\ }\href@noop {} {\bibfield  {journal}
  {\bibinfo  {journal} {Journal of Experimental and Theoretical Physics
  Letters}\ }\textbf {\bibinfo {volume} {78}},\ \bibinfo {pages} {533}
  (\bibinfo {year} {2003})}\BibitemShut {NoStop}%
\bibitem [{\citenamefont {M{\"a}kinen}\ \emph {et~al.}(2023)\citenamefont
  {M{\"a}kinen}, \citenamefont {Autti}, \citenamefont {Heikkinen},
  \citenamefont {Hosio}, \citenamefont {H{\"a}nninen}, \citenamefont {L’vov},
  \citenamefont {Walmsley}, \citenamefont {Zavjalov},\ and\ \citenamefont
  {Eltsov}}]{makinen2023rotating}%
  \BibitemOpen
  \bibfield  {author} {\bibinfo {author} {\bibfnamefont {J.}~\bibnamefont
  {M{\"a}kinen}}, \bibinfo {author} {\bibfnamefont {S.}~\bibnamefont {Autti}},
  \bibinfo {author} {\bibfnamefont {P.}~\bibnamefont {Heikkinen}}, \bibinfo
  {author} {\bibfnamefont {J.}~\bibnamefont {Hosio}}, \bibinfo {author}
  {\bibfnamefont {R.}~\bibnamefont {H{\"a}nninen}}, \bibinfo {author}
  {\bibfnamefont {V.}~\bibnamefont {L’vov}}, \bibinfo {author} {\bibfnamefont
  {P.}~\bibnamefont {Walmsley}}, \bibinfo {author} {\bibfnamefont
  {V.}~\bibnamefont {Zavjalov}},\ and\ \bibinfo {author} {\bibfnamefont
  {V.}~\bibnamefont {Eltsov}},\ }\bibfield  {title} {\bibinfo {title} {Rotating
  quantum wave turbulence},\ }\href@noop {} {\bibfield  {journal} {\bibinfo
  {journal} {Nature physics}\ }\textbf {\bibinfo {volume} {19}},\ \bibinfo
  {pages} {898} (\bibinfo {year} {2023})}\BibitemShut {NoStop}%
\bibitem [{\citenamefont {Hossain}\ \emph {et~al.}(2022)\citenamefont
  {Hossain}, \citenamefont {Kobuszewski}, \citenamefont {Forbes}, \citenamefont
  {Magierski}, \citenamefont {Sekizawa},\ and\ \citenamefont
  {Wlaz\l{}owski}}]{slda-turbulence-2022}%
  \BibitemOpen
  \bibfield  {author} {\bibinfo {author} {\bibfnamefont {K.}~\bibnamefont
  {Hossain}}, \bibinfo {author} {\bibfnamefont {K.}~\bibnamefont
  {Kobuszewski}}, \bibinfo {author} {\bibfnamefont {M.~M.}\ \bibnamefont
  {Forbes}}, \bibinfo {author} {\bibfnamefont {P.}~\bibnamefont {Magierski}},
  \bibinfo {author} {\bibfnamefont {K.}~\bibnamefont {Sekizawa}},\ and\
  \bibinfo {author} {\bibfnamefont {G.}~\bibnamefont {Wlaz\l{}owski}},\
  }\bibfield  {title} {\bibinfo {title} {Rotating quantum turbulence in the
  unitary fermi gas},\ }\href {https://doi.org/10.1103/PhysRevA.105.013304}
  {\bibfield  {journal} {\bibinfo  {journal} {Phys. Rev. A}\ }\textbf {\bibinfo
  {volume} {105}},\ \bibinfo {pages} {013304} (\bibinfo {year}
  {2022})}\BibitemShut {NoStop}%
\bibitem [{\citenamefont {Wlazłowski}\ \emph {et~al.}(2024)\citenamefont
  {Wlazłowski}, \citenamefont {Forbes}, \citenamefont {Sarkar}, \citenamefont
  {Marek},\ and\ \citenamefont {Szpindler}}]{slda-turbulence-2024}%
  \BibitemOpen
  \bibfield  {author} {\bibinfo {author} {\bibfnamefont {G.}~\bibnamefont
  {Wlazłowski}}, \bibinfo {author} {\bibfnamefont {M.~M.}\ \bibnamefont
  {Forbes}}, \bibinfo {author} {\bibfnamefont {S.~R.}\ \bibnamefont {Sarkar}},
  \bibinfo {author} {\bibfnamefont {A.}~\bibnamefont {Marek}},\ and\ \bibinfo
  {author} {\bibfnamefont {M.}~\bibnamefont {Szpindler}},\ }\bibfield  {title}
  {\bibinfo {title} {Fermionic quantum turbulence: Pushing the limits of
  high-performance computing},\ }\bibfield  {journal} {\bibinfo  {journal}
  {PNAS Nexus}\ }\textbf {\bibinfo {volume} {3}},\ \href
  {https://doi.org/10.1093/pnasnexus/pgae160} {10.1093/pnasnexus/pgae160}
  (\bibinfo {year} {2024})\BibitemShut {NoStop}%
\bibitem [{\citenamefont {Donnelly}(1991)}]{donnelly1991quantized}%
  \BibitemOpen
  \bibfield  {author} {\bibinfo {author} {\bibfnamefont {R.~J.}\ \bibnamefont
  {Donnelly}},\ }\href@noop {} {\emph {\bibinfo {title} {Quantized vortices in
  helium II}}},\ Vol.~\bibinfo {volume} {2}\ (\bibinfo  {publisher} {Cambridge
  University Press},\ \bibinfo {year} {1991})\BibitemShut {NoStop}%
\bibitem [{\citenamefont {Simonucci}\ \emph {et~al.}(2019)\citenamefont
  {Simonucci}, \citenamefont {Pieri},\ and\ \citenamefont
  {Strinati}}]{simonucci2019bound}%
  \BibitemOpen
  \bibfield  {author} {\bibinfo {author} {\bibfnamefont {S.}~\bibnamefont
  {Simonucci}}, \bibinfo {author} {\bibfnamefont {P.}~\bibnamefont {Pieri}},\
  and\ \bibinfo {author} {\bibfnamefont {G.~C.}\ \bibnamefont {Strinati}},\
  }\bibfield  {title} {\bibinfo {title} {Bound states in a superfluid vortex: A
  detailed study along the bcs-bec crossover},\ }\href@noop {} {\bibfield
  {journal} {\bibinfo  {journal} {Phys. Rev. B}\ }\textbf {\bibinfo {volume}
  {99}},\ \bibinfo {pages} {134506} (\bibinfo {year} {2019})}\BibitemShut
  {NoStop}%
\bibitem [{\citenamefont {Richaud}\ \emph {et~al.}(2024)\citenamefont
  {Richaud}, \citenamefont {Caldara}, \citenamefont {Capone}, \citenamefont
  {Massignan},\ and\ \citenamefont {Wlaz{\l}owski}}]{richaud2024dynamical}%
  \BibitemOpen
  \bibfield  {author} {\bibinfo {author} {\bibfnamefont {A.}~\bibnamefont
  {Richaud}}, \bibinfo {author} {\bibfnamefont {M.}~\bibnamefont {Caldara}},
  \bibinfo {author} {\bibfnamefont {M.}~\bibnamefont {Capone}}, \bibinfo
  {author} {\bibfnamefont {P.}~\bibnamefont {Massignan}},\ and\ \bibinfo
  {author} {\bibfnamefont {G.}~\bibnamefont {Wlaz{\l}owski}},\ }\bibfield
  {title} {\bibinfo {title} {Dynamical signature of vortex mass in fermi
  superfluids},\ }\href@noop {} {\bibfield  {journal} {\bibinfo  {journal}
  {arXiv preprint arXiv:2410.12417}\ } (\bibinfo {year} {2024})}\BibitemShut
  {NoStop}%
\bibitem [{\citenamefont {Magierski}\ \emph {et~al.}(2024)\citenamefont
  {Magierski}, \citenamefont {Barresi}, \citenamefont {Makowski}, \citenamefont
  {P{\k{e}}cak},\ and\ \citenamefont {Wlaz{\l}owski}}]{magierski2024quantum}%
  \BibitemOpen
  \bibfield  {author} {\bibinfo {author} {\bibfnamefont {P.}~\bibnamefont
  {Magierski}}, \bibinfo {author} {\bibfnamefont {A.}~\bibnamefont {Barresi}},
  \bibinfo {author} {\bibfnamefont {A.}~\bibnamefont {Makowski}}, \bibinfo
  {author} {\bibfnamefont {D.}~\bibnamefont {P{\k{e}}cak}},\ and\ \bibinfo
  {author} {\bibfnamefont {G.}~\bibnamefont {Wlaz{\l}owski}},\ }\bibfield
  {title} {\bibinfo {title} {Quantum vortices in fermionic superfluids: from
  ultracold atoms to neutron stars.},\ }\href@noop {} {\bibfield  {journal}
  {\bibinfo  {journal} {The European Physical Journal A}\ }\textbf {\bibinfo
  {volume} {60}},\ \bibinfo {pages} {186} (\bibinfo {year} {2024})}\BibitemShut
  {NoStop}%
\bibitem [{\citenamefont {Magierski}\ \emph {et~al.}(2022)\citenamefont
  {Magierski}, \citenamefont {Wlaz\l{}owski}, \citenamefont {Makowski},\ and\
  \citenamefont {Kobuszewski}}]{magierski2022reverse}%
  \BibitemOpen
  \bibfield  {author} {\bibinfo {author} {\bibfnamefont {P.}~\bibnamefont
  {Magierski}}, \bibinfo {author} {\bibfnamefont {G.}~\bibnamefont
  {Wlaz\l{}owski}}, \bibinfo {author} {\bibfnamefont {A.}~\bibnamefont
  {Makowski}},\ and\ \bibinfo {author} {\bibfnamefont {K.}~\bibnamefont
  {Kobuszewski}},\ }\bibfield  {title} {\bibinfo {title} {Spin-polarized
  vortices with reversed circulation},\ }\href
  {https://doi.org/10.1103/PhysRevA.106.033322} {\bibfield  {journal} {\bibinfo
   {journal} {Phys. Rev. A}\ }\textbf {\bibinfo {volume} {106}},\ \bibinfo
  {pages} {033322} (\bibinfo {year} {2022})}\BibitemShut {NoStop}%
\bibitem [{\citenamefont {Blatter}\ \emph {et~al.}(1994)\citenamefont
  {Blatter}, \citenamefont {Feigel'man}, \citenamefont {Geshkenbein},
  \citenamefont {Larkin},\ and\ \citenamefont {Vinokur}}]{blatter1994vortices}%
  \BibitemOpen
  \bibfield  {author} {\bibinfo {author} {\bibfnamefont {G.}~\bibnamefont
  {Blatter}}, \bibinfo {author} {\bibfnamefont {M.~V.}\ \bibnamefont
  {Feigel'man}}, \bibinfo {author} {\bibfnamefont {V.~B.}\ \bibnamefont
  {Geshkenbein}}, \bibinfo {author} {\bibfnamefont {A.~I.}\ \bibnamefont
  {Larkin}},\ and\ \bibinfo {author} {\bibfnamefont {V.~M.}\ \bibnamefont
  {Vinokur}},\ }\bibfield  {title} {\bibinfo {title} {Vortices in
  high-temperature superconductors},\ }\href@noop {} {\bibfield  {journal}
  {\bibinfo  {journal} {Reviews of modern physics}\ }\textbf {\bibinfo {volume}
  {66}},\ \bibinfo {pages} {1125} (\bibinfo {year} {1994})}\BibitemShut
  {NoStop}%
\bibitem [{\citenamefont {Fisher}\ \emph {et~al.}(1991)\citenamefont {Fisher},
  \citenamefont {Fisher},\ and\ \citenamefont {Huse}}]{fisher1991thermal}%
  \BibitemOpen
  \bibfield  {author} {\bibinfo {author} {\bibfnamefont {D.~S.}\ \bibnamefont
  {Fisher}}, \bibinfo {author} {\bibfnamefont {M.~P.}\ \bibnamefont {Fisher}},\
  and\ \bibinfo {author} {\bibfnamefont {D.~A.}\ \bibnamefont {Huse}},\
  }\bibfield  {title} {\bibinfo {title} {Thermal fluctuations, quenched
  disorder, phase transitions, and transport in type-ii superconductors},\
  }\href@noop {} {\bibfield  {journal} {\bibinfo  {journal} {Phys. Rev. B}\
  }\textbf {\bibinfo {volume} {43}},\ \bibinfo {pages} {130} (\bibinfo {year}
  {1991})}\BibitemShut {NoStop}%
\bibitem [{\citenamefont {Patel}\ \emph {et~al.}(2020)\citenamefont {Patel},
  \citenamefont {Yan}, \citenamefont {Mukherjee}, \citenamefont {Fletcher},
  \citenamefont {Struck},\ and\ \citenamefont
  {Zwierlein}}]{patel2020universal}%
  \BibitemOpen
  \bibfield  {author} {\bibinfo {author} {\bibfnamefont {P.~B.}\ \bibnamefont
  {Patel}}, \bibinfo {author} {\bibfnamefont {Z.}~\bibnamefont {Yan}}, \bibinfo
  {author} {\bibfnamefont {B.}~\bibnamefont {Mukherjee}}, \bibinfo {author}
  {\bibfnamefont {R.~J.}\ \bibnamefont {Fletcher}}, \bibinfo {author}
  {\bibfnamefont {J.}~\bibnamefont {Struck}},\ and\ \bibinfo {author}
  {\bibfnamefont {M.~W.}\ \bibnamefont {Zwierlein}},\ }\bibfield  {title}
  {\bibinfo {title} {Universal sound diffusion in a strongly interacting fermi
  gas},\ }\href@noop {} {\bibfield  {journal} {\bibinfo  {journal} {Science}\
  }\textbf {\bibinfo {volume} {370}},\ \bibinfo {pages} {1222} (\bibinfo {year}
  {2020})}\BibitemShut {NoStop}%
\bibitem [{\citenamefont {Liu}\ \emph {et~al.}(2024)\citenamefont {Liu},
  \citenamefont {Baggaley}, \citenamefont {Barenghi}, \citenamefont {Wood}
  \emph {et~al.}}]{liu2024vortex}%
  \BibitemOpen
  \bibfield  {author} {\bibinfo {author} {\bibfnamefont {I.}~\bibnamefont
  {Liu}}, \bibinfo {author} {\bibfnamefont {A.~W.}\ \bibnamefont {Baggaley}},
  \bibinfo {author} {\bibfnamefont {C.~F.}\ \bibnamefont {Barenghi}}, \bibinfo
  {author} {\bibfnamefont {T.~S.}\ \bibnamefont {Wood}}, \emph {et~al.},\
  }\bibfield  {title} {\bibinfo {title} {Vortex avalanches and collective
  motion in neutron stars},\ }\href@noop {} {\bibfield  {journal} {\bibinfo
  {journal} {arXiv preprint arXiv:2410.16878}\ } (\bibinfo {year}
  {2024})}\BibitemShut {NoStop}%
\end{thebibliography}

\begin{thebibliography}{76}%
\makeatletter
\providecommand \@ifxundefined [1]{%
 \@ifx{#1\undefined}
}%
\providecommand \@ifnum [1]{%
 \ifnum #1\expandafter \@firstoftwo
 \else \expandafter \@secondoftwo
 \fi
}%
\providecommand \@ifx [1]{%
 \ifx #1\expandafter \@firstoftwo
 \else \expandafter \@secondoftwo
 \fi
}%
\providecommand \natexlab [1]{#1}%
\providecommand \enquote  [1]{``#1''}%
\providecommand \bibnamefont  [1]{#1}%
\providecommand \bibfnamefont [1]{#1}%
\providecommand \citenamefont [1]{#1}%
\providecommand \href@noop [0]{\@secondoftwo}%
\providecommand \href [0]{\begingroup \@sanitize@url \@href}%
\providecommand \@href[1]{\@@startlink{#1}\@@href}%
\providecommand \@@href[1]{\endgroup#1\@@endlink}%
\providecommand \@sanitize@url [0]{\catcode `\\12\catcode `\$12\catcode
  `\&12\catcode `\#12\catcode `\^12\catcode `\_12\catcode `\%12\relax}%
\providecommand \@@startlink[1]{}%
\providecommand \@@endlink[0]{}%
\providecommand \url  [0]{\begingroup\@sanitize@url \@url }%
\providecommand \@url [1]{\endgroup\@href {#1}{\urlprefix }}%
\providecommand \urlprefix  [0]{URL }%
\providecommand \Eprint [0]{\href }%
\providecommand \doibase [0]{https://doi.org/}%
\providecommand \selectlanguage [0]{\@gobble}%
\providecommand \bibinfo  [0]{\@secondoftwo}%
\providecommand \bibfield  [0]{\@secondoftwo}%
\providecommand \translation [1]{[#1]}%
\providecommand \BibitemOpen [0]{}%
\providecommand \bibitemStop [0]{}%
\providecommand \bibitemNoStop [0]{.\EOS\space}%
\providecommand \EOS [0]{\spacefactor3000\relax}%
\providecommand \BibitemShut  [1]{\csname bibitem#1\endcsname}%
\let\auto@bib@innerbib\@empty
\bibitem [65]{zurn2013precise}%
  \BibitemOpen
  \bibfield  {author} {\bibinfo {author} {\bibfnamefont {G.}~\bibnamefont
  {Z{\"u}rn}}, \bibinfo {author} {\bibfnamefont {T.}~\bibnamefont {Lompe}},
  \bibinfo {author} {\bibfnamefont {A.~N.}\ \bibnamefont {Wenz}}, \bibinfo
  {author} {\bibfnamefont {S.}~\bibnamefont {Jochim}}, \bibinfo {author}
  {\bibfnamefont {P.}~\bibnamefont {Julienne}},\ and\ \bibinfo {author}
  {\bibfnamefont {J.}~\bibnamefont {Hutson}},\ }\bibfield  {title} {\bibinfo
  {title} {Precise characterization of li 6 feshbach resonances using
  trap-sideband-resolved rf spectroscopy of weakly bound
  molecules},\ }\href@noop {} {\bibfield  {journal} {\bibinfo  {journal} {Phys.
  Rev. Lett.}\ }\textbf {\bibinfo {volume} {110}},\ \bibinfo {pages} {135301}
  (\bibinfo {year} {2013})}\BibitemShut {NoStop}%
\bibitem [66]{hernandez2024connecting}%
  \BibitemOpen
  \bibfield  {author} {\bibinfo {author} {\bibfnamefont {D.}~\bibnamefont
  {Hernández-Rajkov}}, \bibinfo {author} {\bibfnamefont {N.}~\bibnamefont
  {Grani}}, \bibinfo {author} {\bibfnamefont {F.}~\bibnamefont {Scazza}},
  \bibinfo {author} {\bibfnamefont {G.}~\bibnamefont {Del~Pace}}, \bibinfo
  {author} {\bibfnamefont {W.~J.}\ \bibnamefont {Kwon}}, \bibinfo {author}
  {\bibfnamefont {M.}~\bibnamefont {Inguscio}}, \bibinfo {author}
  {\bibfnamefont {K.}~\bibnamefont {Xhani}}, \bibinfo {author} {\bibfnamefont
  {C.}~\bibnamefont {Fort}}, \bibinfo {author} {\bibfnamefont {M.}~\bibnamefont
  {Modugno}}, \bibinfo {author} {\bibfnamefont {F.}~\bibnamefont {Marino}},\
  and\ \bibinfo {author} {\bibfnamefont {G.}~\bibnamefont {Roati}},\ }\bibfield
   {title} {\bibinfo {title} {Connecting shear flow and vortex array
  instabilities in annular atomic superfluids},\ }\href
  {https://doi.org/10.1038/s41567-024-02466-4} {\bibfield  {journal} {\bibinfo
  {journal} {Nature Physics}\ }\textbf {\bibinfo {volume} {20}},\ \bibinfo
  {pages} {939–944} (\bibinfo {year} {2024})}\BibitemShut {NoStop}%
\bibitem [67]{del2021tunneling}%
  \BibitemOpen
  \bibfield  {author} {\bibinfo {author} {\bibfnamefont {G.}~\bibnamefont
  {Del~Pace}}, \bibinfo {author} {\bibfnamefont {W.}~\bibnamefont {Kwon}},
  \bibinfo {author} {\bibfnamefont {M.}~\bibnamefont {Zaccanti}}, \bibinfo
  {author} {\bibfnamefont {G.}~\bibnamefont {Roati}},\ and\ \bibinfo {author}
  {\bibfnamefont {F.}~\bibnamefont {Scazza}},\ }\bibfield  {title} {\bibinfo
  {title} {Tunneling transport of unitary fermions across the superfluid
  transition},\ }\href@noop {} {\bibfield  {journal} {\bibinfo  {journal}
  {Physical Review Letters}\ }\textbf {\bibinfo {volume} {126}},\ \bibinfo
  {pages} {055301} (\bibinfo {year} {2021})}\BibitemShut {NoStop}%
\bibitem [68]{ku2012revealing}%
  \BibitemOpen
  \bibfield  {author} {\bibinfo {author} {\bibfnamefont {M.~J.}\ \bibnamefont
  {Ku}} \emph {et~al.},\ }\bibfield  {title} {\bibinfo {title} {Revealing the
  superfluid lambda transition in the universal thermodynamics of a unitary
  fermi gas},\ }\href {https://www.science.org/doi/10.1126/science.1214987}
  {\bibfield  {journal} {\bibinfo  {journal} {Science}\ }\textbf {\bibinfo
  {volume} {335}},\ \bibinfo {pages} {563} (\bibinfo {year}
  {2012})}\BibitemShut {NoStop}%
\bibitem [69]{kwon2020strongly}%
  \BibitemOpen
  \bibfield  {author} {\bibinfo {author} {\bibfnamefont {W.}~\bibnamefont
  {Kwon}}, \bibinfo {author} {\bibfnamefont {G.}~\bibnamefont {Del~Pace}},
  \bibinfo {author} {\bibfnamefont {R.}~\bibnamefont {Panza}}, \bibinfo
  {author} {\bibfnamefont {M.}~\bibnamefont {Inguscio}}, \bibinfo {author}
  {\bibfnamefont {W.}~\bibnamefont {Zwerger}}, \bibinfo {author} {\bibfnamefont
  {M.}~\bibnamefont {Zaccanti}}, \bibinfo {author} {\bibfnamefont
  {F.}~\bibnamefont {Scazza}},\ and\ \bibinfo {author} {\bibfnamefont
  {G.}~\bibnamefont {Roati}},\ }\bibfield  {title} {\bibinfo {title} {Strongly
  correlated superfluid order parameters from dc josephson supercurrents},\
  }\href@noop {} {\bibfield  {journal} {\bibinfo  {journal} {Science}\ }\textbf
  {\bibinfo {volume} {369}},\ \bibinfo {pages} {84} (\bibinfo {year}
  {2020})}\BibitemShut {NoStop}%
\bibitem [70]{Sonin_magnusForce}%
  \BibitemOpen
  \bibfield  {author} {\bibinfo {author} {\bibfnamefont {E.~B.}\ \bibnamefont
  {Sonin}},\ }\bibfield  {title} {\bibinfo {title} {Magnus force in superfluids
  and superconductors},\ }\href {https://doi.org/10.1103/PhysRevB.55.485}
  {\bibfield  {journal} {\bibinfo  {journal} {Phys. Rev. B}\ }\textbf {\bibinfo
  {volume} {55}},\ \bibinfo {pages} {485} (\bibinfo {year} {1997})}\BibitemShut
  {NoStop}%
\bibitem [71]{Imaginary_vortex_Disk}%
  \BibitemOpen
  \bibfield  {author} {\bibinfo {author} {\bibfnamefont {J.-P.}\ \bibnamefont
  {Martikainen}} \emph {et~al.},\ }\bibfield  {title} {\bibinfo {title}
  {Generation and evolution of vortex-antivortex pairs in bose-einstein
  condensates},\ }\href {https://doi.org/10.1103/PhysRevA.64.063602} {\bibfield
   {journal} {\bibinfo  {journal} {Phys. Rev. A}\ }\textbf {\bibinfo {volume}
  {64}},\ \bibinfo {pages} {063602} (\bibinfo {year} {2001})}\BibitemShut
  {NoStop}%
\bibitem [72]{galantucci2020new}%
  \BibitemOpen
  \bibfield  {author} {\bibinfo {author} {\bibfnamefont {L.}~\bibnamefont
  {Galantucci}}, \bibinfo {author} {\bibfnamefont {A.~W.}\ \bibnamefont
  {Baggaley}}, \bibinfo {author} {\bibfnamefont {C.~F.}\ \bibnamefont
  {Barenghi}},\ and\ \bibinfo {author} {\bibfnamefont {G.}~\bibnamefont
  {Krstulovic}},\ }\bibfield  {title} {\bibinfo {title} {A new self-consistent
  approach of quantum turbulence in superfluid helium},\ }\href@noop {}
  {\bibfield  {journal} {\bibinfo  {journal} {The European Physical Journal
  Plus}\ }\textbf {\bibinfo {volume} {135}},\ \bibinfo {pages} {1} (\bibinfo
  {year} {2020})}\BibitemShut {NoStop}%
\bibitem [73]{kopnin1995spectral}%
  \BibitemOpen
  \bibfield  {author} {\bibinfo {author} {\bibfnamefont {N.}~\bibnamefont
  {Kopnin}}, \bibinfo {author} {\bibfnamefont {G.}~\bibnamefont {Volovik}},\
  and\ \bibinfo {author} {\bibfnamefont {{\"U}.}~\bibnamefont {Parts}},\
  }\bibfield  {title} {\bibinfo {title} {Spectral flow in vortex dynamics of
  3he-b and superconductors},\ }\href@noop {} {\bibfield  {journal} {\bibinfo
  {journal} {Europhysics Letters}\ }\textbf {\bibinfo {volume} {32}},\ \bibinfo
  {pages} {651} (\bibinfo {year} {1995})}\BibitemShut {NoStop}%
\bibitem [74]{Goodman1981}%
  \BibitemOpen
  \bibfield  {author} {\bibinfo {author} {\bibfnamefont {A.~L.}\ \bibnamefont
  {Goodman}},\ }\bibfield  {title} {\bibinfo {title} {Finite-temperature {HFB}
  theory},\ }\href {https://doi.org/10.1016/0375-9474(81)90557-1} {\bibfield
  {journal} {\bibinfo  {journal} {Nucl. Phys. A}\ }\textbf {\bibinfo {volume}
  {352}},\ \bibinfo {pages} {30} (\bibinfo {year} {1981})}\BibitemShut
  {NoStop}%
\bibitem [75]{Bulgac2012}%
  \BibitemOpen
  \bibfield  {author} {\bibinfo {author} {\bibfnamefont {A.}~\bibnamefont
  {Bulgac}}, \bibinfo {author} {\bibfnamefont {M.~M.}\ \bibnamefont {Forbes}},\
  and\ \bibinfo {author} {\bibfnamefont {P.}~\bibnamefont {Magierski}},\
  }\bibfield  {title} {\bibinfo {title} {The {{Unitary Fermi Gas}}: From
  {{Monte Carlo}} to {{Density Functionals}}},\ }in\ \href
  {https://doi.org/10.1007/978-3-642-21978-8_9} {\emph {\bibinfo {booktitle}
  {The {{BCS}}-{{BEC Crossover}} and the {{Unitary Fermi Gas}}}}},\ \bibinfo
  {series and number} {Lecture {{Notes}} in {{Physics}}},\ \bibinfo {editor}
  {edited by\ \bibinfo {editor} {\bibfnamefont {W.}~\bibnamefont {Zwerger}}}\
  (\bibinfo  {publisher} {{Springer}},\ \bibinfo {address} {{Berlin,
  Heidelberg}},\ \bibinfo {year} {2012})\ pp.\ \bibinfo {pages}
  {305--373}\BibitemShut {NoStop}%
\bibitem [76]{WSLDAToolkit}%
  \BibitemOpen
  \href {https://wslda.fizyka.pw.edu.pl/} {\bibinfo {title} {{W-SLDA
  Toolkit}}},\ \bibinfo {howpublished}
  {\url{https://wslda.fizyka.pw.edu.pl/}}\BibitemShut {NoStop}%
\end{thebibliography}

\begin{thebibliography}{12}%
\makeatletter
\providecommand \@ifxundefined [1]{%
 \@ifx{#1\undefined}
}%
\providecommand \@ifnum [1]{%
 \ifnum #1\expandafter \@firstoftwo
 \else \expandafter \@secondoftwo
 \fi
}%
\providecommand \@ifx [1]{%
 \ifx #1\expandafter \@firstoftwo
 \else \expandafter \@secondoftwo
 \fi
}%
\providecommand \natexlab [1]{#1}%
\providecommand \enquote  [1]{``#1''}%
\providecommand \bibnamefont  [1]{#1}%
\providecommand \bibfnamefont [1]{#1}%
\providecommand \citenamefont [1]{#1}%
\providecommand \href@noop [0]{\@secondoftwo}%
\providecommand \href [0]{\begingroup \@sanitize@url \@href}%
\providecommand \@href[1]{\@@startlink{#1}\@@href}%
\providecommand \@@href[1]{\endgroup#1\@@endlink}%
\providecommand \@sanitize@url [0]{\catcode `\\12\catcode `\$12\catcode
  `\&12\catcode `\#12\catcode `\^12\catcode `\_12\catcode `\%12\relax}%
\providecommand \@@startlink[1]{}%
\providecommand \@@endlink[0]{}%
\providecommand \url  [0]{\begingroup\@sanitize@url \@url }%
\providecommand \@url [1]{\endgroup\@href {#1}{\urlprefix }}%
\providecommand \urlprefix  [0]{URL }%
\providecommand \Eprint [0]{\href }%
\providecommand \doibase [0]{https://doi.org/}%
\providecommand \selectlanguage [0]{\@gobble}%
\providecommand \bibinfo  [0]{\@secondoftwo}%
\providecommand \bibfield  [0]{\@secondoftwo}%
\providecommand \translation [1]{[#1]}%
\providecommand \BibitemOpen [0]{}%
\providecommand \bibitemStop [0]{}%
\providecommand \bibitemNoStop [0]{.\EOS\space}%
\providecommand \EOS [0]{\spacefactor3000\relax}%
\providecommand \BibitemShut  [1]{\csname bibitem#1\endcsname}%
\let\auto@bib@innerbib\@empty
\bibitem [77]{Haussmann1993}%
  \BibitemOpen
  \bibfield  {author} {\bibinfo {author} {\bibfnamefont {R.}~\bibnamefont
  {Haussmann}},\ }\bibfield  {title} {\bibinfo {title} {{Crossover from BCS
  superconductivity to Bose-Einstein condensation: A self-consistent theory}},\
  }\href@noop {} {\bibfield  {journal} {\bibinfo  {journal} {Z. Phys. B}\
  }\textbf {\bibinfo {volume} {91}},\ \bibinfo {pages} {291} (\bibinfo {year}
  {1993})}\BibitemShut {NoStop}%
\bibitem [78]{Haussmann1994}%
  \BibitemOpen
  \bibfield  {author} {\bibinfo {author} {\bibfnamefont {R.}~\bibnamefont
  {Haussmann}},\ }\bibfield  {title} {\bibinfo {title} {Properties of a fermi
  liquid at the superfluid transition in the crossover region between bcs
  superconductivity and bose-einstein condensation},\ }\href
  {https://doi.org/10.1103/PhysRevB.49.12975} {\bibfield  {journal} {\bibinfo
  {journal} {Phys. Rev. B}\ }\textbf {\bibinfo {volume} {49}},\ \bibinfo
  {pages} {12975} (\bibinfo {year} {1994})}\BibitemShut {NoStop}%
\bibitem [79]{Haussmann2008}%
  \BibitemOpen
  \bibfield  {author} {\bibinfo {author} {\bibfnamefont {R.}~\bibnamefont
  {Haussmann}}\ and\ \bibinfo {author} {\bibfnamefont {W.}~\bibnamefont
  {Zwerger}},\ }\bibfield  {title} {\bibinfo {title} {Thermodynamics of a
  trapped unitary fermi gas},\ }\href
  {https://doi.org/10.1103/PhysRevA.78.063602} {\bibfield  {journal} {\bibinfo
  {journal} {Phys. Rev. A}\ }\textbf {\bibinfo {volume} {78}},\ \bibinfo
  {pages} {063602} (\bibinfo {year} {2008})}\BibitemShut {NoStop}%
\bibitem [80]{Pini2020}%
  \BibitemOpen
  \bibfield  {author} {\bibinfo {author} {\bibfnamefont {M.}~\bibnamefont
  {Pini}}, \bibinfo {author} {\bibfnamefont {P.}~\bibnamefont {Pieri}},
  \bibinfo {author} {\bibfnamefont {M.}~\bibnamefont {Jäger}}, \bibinfo
  {author} {\bibfnamefont {J.~H.}\ \bibnamefont {Denschlag}},\ and\ \bibinfo
  {author} {\bibfnamefont {G.~C.}\ \bibnamefont {Strinati}},\ }\bibfield
  {title} {\bibinfo {title} {Pair correlations in the normal phase of an
  attractive fermi gas},\ }\href {https://doi.org/10.1088/1367-2630/ab9ee3}
  {\bibfield  {journal} {\bibinfo  {journal} {New Journal of Physics}\ }\textbf
  {\bibinfo {volume} {22}},\ \bibinfo {pages} {083008} (\bibinfo {year}
  {2020})}\BibitemShut {NoStop}%
\bibitem [81]{Pini2021}%
  \BibitemOpen
  \bibfield  {author} {\bibinfo {author} {\bibfnamefont {M.}~\bibnamefont
  {Pini}}, \bibinfo {author} {\bibfnamefont {P.}~\bibnamefont {Pieri}},
  \bibinfo {author} {\bibfnamefont {R.}~\bibnamefont {Grimm}},\ and\ \bibinfo
  {author} {\bibfnamefont {G.~C.}\ \bibnamefont {Strinati}},\ }\bibfield
  {title} {\bibinfo {title} {Beyond-mean-field description of a trapped unitary
  fermi gas with mass and population imbalance},\ }\href
  {https://doi.org/10.1103/PhysRevA.103.023314} {\bibfield  {journal} {\bibinfo
   {journal} {Phys. Rev. A}\ }\textbf {\bibinfo {volume} {103}},\ \bibinfo
  {pages} {023314} (\bibinfo {year} {2021})}\BibitemShut {NoStop}%
\bibitem [82]{Thouless1960}%
  \BibitemOpen
  \bibfield  {author} {\bibinfo {author} {\bibfnamefont {D.~J.}\ \bibnamefont
  {Thouless}},\ }\bibfield  {title} {\bibinfo {title} {Perturbation theory in
  statistical mechanics and the theory of superconductivity},\ }\href@noop {}
  {\bibfield  {journal} {\bibinfo  {journal} {Ann. Phys.}\ }\textbf {\bibinfo
  {volume} {10}},\ \bibinfo {pages} {553} (\bibinfo {year} {1960})}\BibitemShut
  {NoStop}%
\end{thebibliography}
\end{document}